# Transformer-refined quantum sampling for strongly correlated electronic structure

Xiongzhi Zeng[1†], Ming Gong[2,3,4†], Bowen Kan[1†] , Yi Fan[1], Huan Ma[1], Jianbin Cai[2,3,4], Yancheng Liu[2,3,4], Naibin Zhou[2,3,4], Tao Jiang[2,3,4], Shaojun Guo[2,3,4], Zhijie Fan[2,3,4], Zongkang Zhang[2,3,4], Yuan Li[2,3,4], Sirui Cao[2,3,4], Kai Yan[2,3,4], Xiaobo Zhu[2,3,4], Yi Luo[2,4], Honghui Shang[1*], Zhenyu Li[1,4*], Jian-Wei Pan[2,3,4*]and Jinlong Yang[1,4*]

[1]State Key Laboratory of Precision and Intelligent Chemistry, University of Science and Technology of China, Hefei 230026, China

[2]Hefei National Research Center for Physical Sciences at the Microscale and School of Physical Sciences, University of Science and Technology of China, Hefei 230026, China

[3]Shanghai Research Center for Quantum Science and CAS Center for Excellence in Quantum Information and Quantum Physics, University of Science and Technology of China, Shanghai 201315, China

[4]Hefei National Laboratory, University of Science and Technology of China, Hefei 230088, China

[†]These authors contribute equally to this work.

[*]To whom correspondence should be addressed:

shanghui.ustc@gmail.com; zyli@ustc.edu.cn;

pan@ustc.edu.cn; jlyang@ustc.edu.cn

## Abstract:

Although quantum computing offers a promising solution for strongly correlated system simulation, existing algorithms face significant bottlenecks on current noisy intermediate-scale quantum (NISQ) devices. Here, we introduce QiankunNet-QSCI, a hybrid quantum-classical framework that addresses this challenge by combining efficient quantum-sampling with a transformer neural network. An efficient unitary selected configuration Interaction (USCI) ansatz especially designed for quantum sampling is proposed to identify the most chemically significant electronic configurations on the *Zuchongzhi 3.1* quantum processor. Subsequently, the transformer model *QiankunNet* learns from these sparse yet critical quantum data to infer and reconstruct the complete electronic wavefunction with high fidelity. Simulation of the challenging 40-qubit $[Fe_2S_2(SCH_3)_4]^{2-}$ ([2Fe-2S]) ferredoxin active center achieves chemical accuracy. Simulation of the nitrogenase P-cluster in a 114-electron 73-orbital active space also reaches ~12 milli-Hartree-level agreement with the best density matrix renormalization group (DMRG) result. QiankunNet-QSCI thus offers a practical route to accurate quantum-assisted electronic structure calculations on current devices.

Accurately predicting the electronic structure of molecules and materials is a central challenge in chemistry and materials science[1,2]. Strongly correlated electron systems – such as transition metal clusters – are especially difficult because there are many near-degenerated electron configurations. A broad spectrum of classical methods, from density functional theory (DFT) to advanced wavefunction approaches[3-4], have been developed to obtain approximate solutions of the electronic structure problem. DFT is broadly successful but most of its functionals suffer from well-documented delocalization and static-correlation errors on multi-reference problems, limiting its reliability for strongly correlated systems[4-6]. Among wavefunction methods, tensor-network approaches such as the density-matrix renormalization group (DMRG) are powerful and accurate in spaces compatible with area-law entanglement (e.g., quasi-one-dimensional orbital topologies)[7]. However, in other cases, the bond dimension (and cost) must hugely grow to capture electron correlations[7,8]. Therefore, even if it provides state-of-the-art results for some strongly correlated systems, DMRG is not considered to be a method generally scalable for large systems.

Quantum computing is of the potential to revolutionize quantum chemistry by efficiently representing many-body wavefunctions through quantum superposition and entanglement[9,10]. However, even though quantum algorithms such as quantum phase estimation can, in principle, extract molecular ground-state energies in time polynomial in the system size[11,12]—an exponential improvement over exact classical approaches such as full configuration interaction (FCI)—the enormous logical-gate counts, coupled with short coherence times and high error rates of today's noisy intermediate-scale quantum (NISQ) hardware, render a fault-tolerant implementation out of reach for the foreseeable future[13]. To bridge the gap, hybrid quantum-classical strategies have emerged that offload parts of the computation on quantum devices by leveraging classical computing[14]. The most notable frameworks—variational quantum eigensolver (VQE)[15] and quantum selected configuration interaction (QSCI)[16,17]—show great potential in solving electronic structure problems. Despite their promise, both approaches face well-documented bottlenecks. In VQE, it is difficult to design ansatz that is simultaneously expressive, hardware-efficient, and symmetry-preserving[18,19]; even when such a circuit is identified, noisy gradients, barren-plateau landscapes[20], and an exponentially growing number of measurements complicate parameter optimization[21]. For QSCI, the quantum sampling of determinants often produces large volumes of redundant or low-weight configurations, wasting valuable quantum resources and slowing convergence[22]. Researchers have started combining quantum data with machine learning to address current quantum computing challenges[23,24,25]. While promising, these approaches typically do not prioritize chemically meaningful determinants at the data acquisition stage and only weakly encode quantum-chemical structure—exact electron-number and spin symmetries, point-group constraints, and CI-space sparsity—when learning from noisy, sparsely sampled data.

In this study, we present QiankunNet-QSCI, a hybrid quantum–classical framework that couples focused quantum sampling with physics-informed transformer refinement to deliver chemical accuracy on strongly correlated molecules. Rather than as a generic variational engine, the quantum processor is used in the QSCI stage as a focused determinant sampler, where a compact Unitary Selected Configuration Interaction (USCI) ansatz specially designed for this purpose is proposed to enrich the chemically dominant determinant sector on hardware. Then, the transformer neural network *QiankunNet*[26] denoises, completes and variationally refines the sampled wavefunction. Such a two-stage design channels both quantum and classical resources toward the most chemically relevant sector of Hilbert space, markedly accelerating convergence while eliminating unnecessary sampling overhead. We first validate our framework on small molecules including $H_{10}$ ring[27], then demonstrate chemical accuracy relative to DMRG for the 40-qubit $[Fe_2S_2(SCH_3)_4]^{2-}$ ferredoxin cluster[28] on the *Zuchongzhi* 3.1 quantum processor[29], which to the best of our knowledge is among the first demonstrations to reach this accuracy at this scale with a quantum processor. We also successfully applied this protocol to the nitrogenase P-cluster[30,31] in a (114e,73o) complete active space (CAS). This represents one of the largest quantum-chemistry active-space calculations demonstrated on a superconducting quantum processor, achieving energy within ~18 mHa of the DMRG extrapolation reference. These results establish QiankunNet-QSCI as a practical pathway for quantum-hardware-assisted computational chemistry, demonstrating that current NISQ hardware can tackle chemically relevant problems when paired with intelligent classical processing.

**Architecture of QiankunNet-QSCI:** As illustrated in Fig. 1, QiankunNet-QSCI operates through a two-stage process that maximizes the utility of both quantum and classical resources (details are given in Methods/ Supplementary Materials). In the quantum step, we design a USCI ansatz to prepare a symmetry-consistent, prescreened determinant set (configurations) via quantum circuit guided importance sampling of the configuration space, and then perform subspace diagonalization to obtain initial amplitudes. The classical learning step feeds the filtered determinants to *QiankunNet*—a transformer with amplitude/phase-decoupled heads—to reweight, denoise, and variationally refine coefficients, correcting hardware-induced biases and yielding unbiased amplitudes for property evaluation. *QiankunNet*[26] adopts a decoder-only transformer architecture that processes each Slater determinant's occupation string as an input sequence, using multi-head self-attention layers to capture correlations between orbitals. By training on the set of important determinants obtained from the quantum step and their corresponding coefficients, *QiankunNet* learns an improved approximation to the ground-state wavefunction via a variational refinement.

QiankunNet-QSCI achieves its breakthrough performance through three integrated advances:

(i) Quantum hardware capabilities: We leverage the *Zuchongzhi 3.1* superconducting quantum processor, a 105-qubit system refined from *Zuchongzhi 3.0*[29]. Its one-to-four tunable coupling architecture enables dynamic optimization of physical circuit layouts based on algorithmic requirements. The processor delivers exceptional fidelity metrics critical for quantum chemistry: single-qubit gate errors of ~0.1%, two-qubit gate errors of ~0.6%, and readout errors of ~0.6%. Combined with an integrated control system supporting ~1.4 kHz effective sampling rates, we collected 100 million high-fidelity samples in approximately 20 hours—a throughput essential for capturing the complex correlation patterns in strongly correlated molecules. This hardware performance represents a significant advance over previous NISQ devices, providing the foundation for chemically accurate quantum-hardware experiments at the 40-qubit scale.

(ii) Efficient USCI ansatz: For an ansatz designed for quantum sampling, more completeness can be traded off for compactness. The USCI ansatz proposed in this study is designed following such a principle and thus extremely compact. For example, for the [2Fe-2S] cluster, which typically leads to circuits with a depth >250 with thousands of two-qubit gates, USCI circuit is of depth 41 with only 795 single-qubit and 193 two-qubit gates (49 parameters). Across a chemically diverse set—spanning different elements, spin states, and correlation patterns—USCI consistently requires only 10-15 determinants to construct and fewer than 1,300 two-qubit gates in total (Table S2). The efficiency of USCI guarantees that even NISQ devices deliver informative samples rather than noise-dominated tails.

(iii) Synergy of quantum computation and artificial intelligence: Instead of using it as a variational eigenvalue solver which is significantly limited by noise, we use quantum computer only to grasp the main electron correlation pattern. The quantum information obtained is then used to guide QiankunNet to find the ground state, which makes it possible to accurately calculate electronic structure for systems otherwise beyond reach with random initialization of QiankunNet.

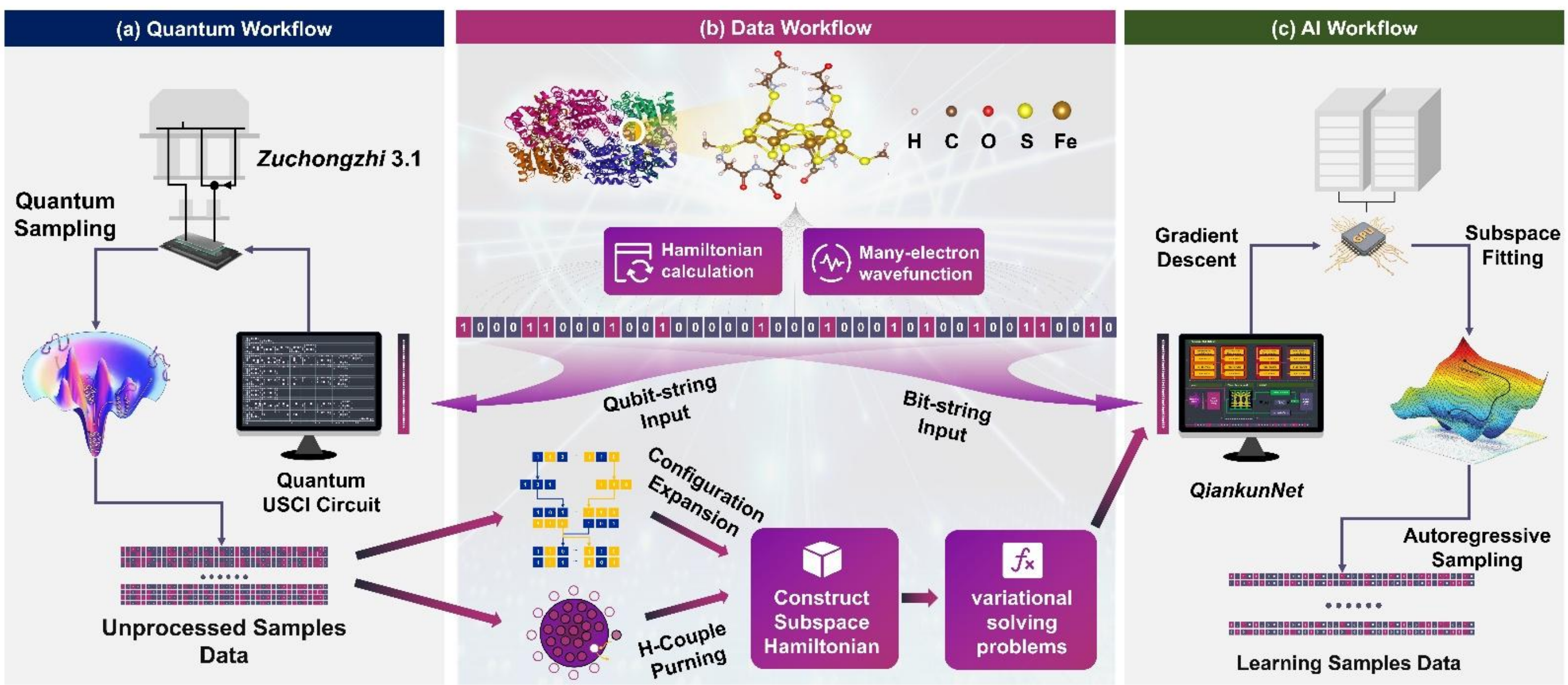

**Fig. 1 QiankunNet-QSCI workflow.** (a) Quantum workflow, in which a compact USCI ansatz is executed on quantum hardware to generate focused determinant samples. (b) Data workflow, where raw bitstrings are symmetry filtered, expanded, and assembled into a constrained subspace Hamiltonian. (c) AI workflow, in which QiankunNet is trained to denoise, complete, and refine the sampled wavefunction.

**Quantum hardware validation on $H_{10}$ ring.** Using some small molecules including $N_2$, $O_2$, $Cr_2$, $Cu_2O_2^{2+}$, cyclobutadiene ($C_4H_4$)[32-36] as examples, based on classical simulations, we confirm that QiankunNet-QSCI can efficiently predict the ground state with a high accuracy (see Supplementary Materials). Then, we validate our approach through proof-of-concept experiments on superconducting quantum processors. Using *Zuchongzhi 3.1*(Table 1), we executed quantum circuits for an $H_{10}$ ring consisting of ten hydrogen atoms uniformly arranged on a circle with a radius of 1.0 Å[27] —a 20-qubit problem with an FCI space of 63,504 determinants, of which approximately 31,388 have non-negligible amplitudes.

On NISQ devices, the quality of the ansatz design critically determines whether the chemically informative quantum signal can be effectively preserved in the presence of noise. We first validate the effectiveness of our quantum sampling protocol by comparing our tailored USCI ansatz with the Local Unitary Coupled Cluster with Jastrow (LUCJ) ansatz previously used in QSCI[37] on this benchmark system of $H_{10}$. In experiment, each circuit was executed on the quantum device to generate one million measurement samples. The results revealed a striking performance difference. The LUCJ ansatz produced ~635,000 unique bitstrings, but only 38,608 satisfied the physical constraint of 10 electrons. Moreover, the measured bitstrings were broadly distributed across Hilbert space—none of the ten most frequent outputs were corresponded to the true dominant configuration (see Fig.S8 in Supplementary Materials), which demonstrates that noise indeed significantly degrade the sampling precision.

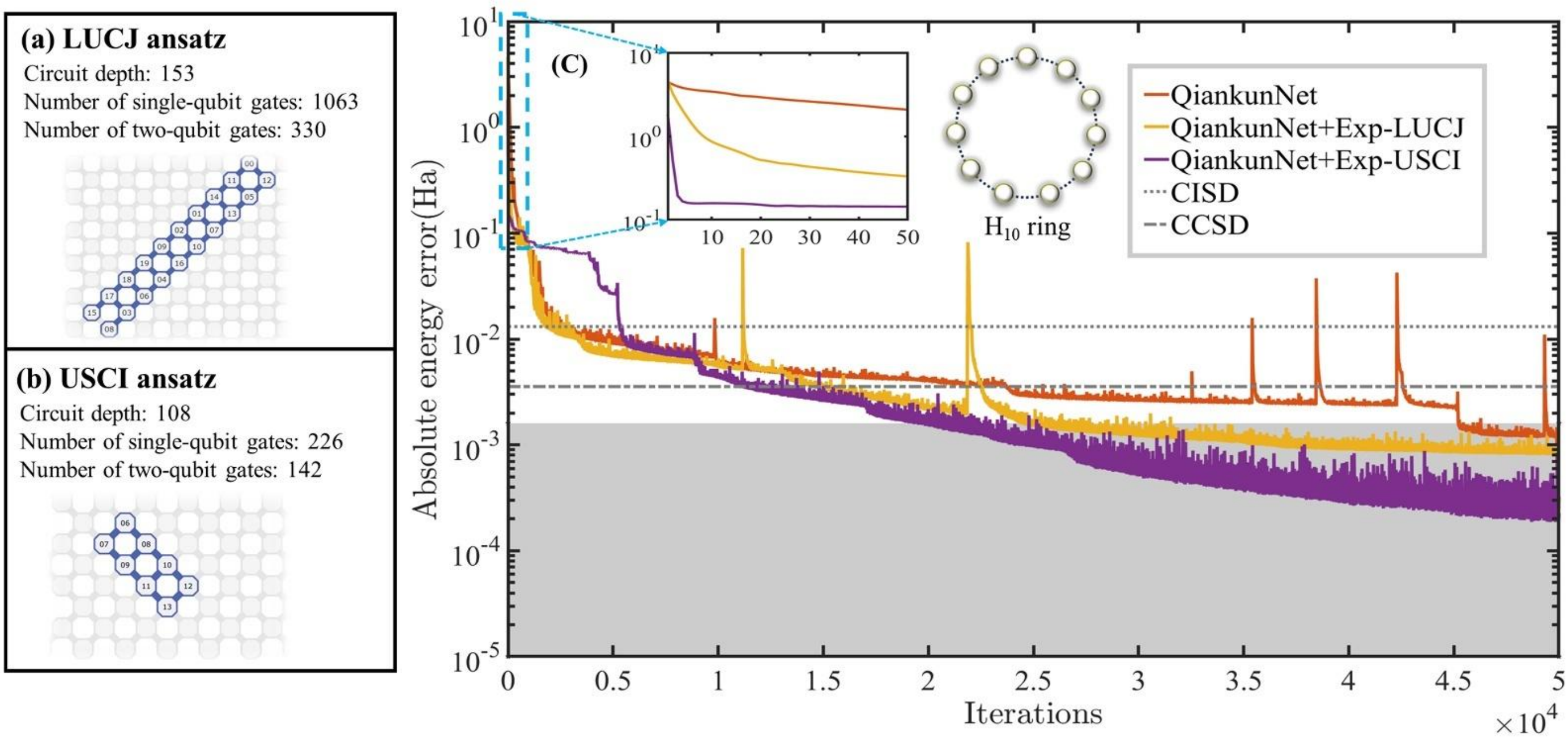

**Fig. 2 Quantum Hardware Validation on $H_{10}$ Ring System.** (a) Layout of the 20 qubits on *Zuchongzhi* 3.1 with the LUCJ entangler path highlighted in blue. (b) Qubit sub-lattice used for the 8-qubit USCI cluster highlighted in blue. (c) Absolute energy error versus iteration for QiankunNet with random initialization (orange), QiankunNet initialized with LUCJ (yellow), and QiankunNet initialized with USCI (purple). Energy errors from classical methods CISD (dotted line) and CCSD (dashed line) are provided for comparison. The shaded gray region marks the chemical accuracy threshold ($1.6\times10^{-3}$ Ha). The inset provides an enlarged view of the early iterations and the structure of $H_{10}$ ring.

In USCI ansatz construction, we select seven configurations to start with which only involve 8 of the 29 orbitals. The resulting circuit is of about half depth compared to the LUCJ circuit. Therefore, the USCI ansatz produce a much more concentrated measurement distribution on the same problem. From 1,000,000 samples, we obtained only 256 unique bitstrings, with 36 maintaining the correct particle number. This large reduction in unique outcomes indicates that USCI is more noise-resilient and it directs probability mass toward a small set of physically relevant configurations. The sampled determinants also included the known dominant configuration of the $H_{10}$ ring (see Fig.S8 in Supplementary Materials), showing that the ansatz effectively pre-selects chemically important sectors. Rather than post-selecting valid configurations from hundreds of thousands of possibilities, this ansatz enables the quantum hardware directly produce a focused set of physically meaningful states.

Such a sampling effectiveness can be directly translated into algorithmic performance. When using these USCI-generated samples to train *QiankunNet*, the network immediately operates with a near-optimal determinant basis. Fig. 2 demonstrates this advantage through error convergence curves: QiankunNet+USCI (purple) reaches the $1.6\times10^{-3}$ Hartree accuracy threshold fast, while QiankunNet+LUCJ (yellow) converges more slowly, and QiankunNet starting from random initialization without quantum sampling (orange) exhibits slow, oscillatory convergence. The inset magnifies the critical initial iterations, where QSCI enables error reduction by orders of magnitude within just a few steps. These data demonstrate that both components—the physics-informed ansatz and neural network refinement—are essential for QiankunNet-QSCI's rapid convergence.

**Reach chemical accuracy for 40-qubit $[Fe_2S_2(SCH_3)_4]^{2-}$ cluster.** The $[Fe_2S_2(SCH_3)_4]^{2-}$ ferredoxin cluster is a stringent benchmark for quantum-chemistry algorithms because its electronic structure exhibits strong correlation and multiple low-lying spin states[28]. In this study, its geometry is adopted from the experimental structure reported in Ref 38. Its electronic structure is described by a 30-electron, 20-orbital active space (40 spin-orbitals, hence 40 qubits) with TZP-DKH basis set. In Fig. 3, we summarize our comprehensive benchmarks comparing QiankunNet-QSCI against state-of-the-art classical and quantum-enabled methods. For the quantum sampling, we deployed the USCI ansatz on *Zuchongzhi 3.1* (Table 2) utilizing 40 qubits with a circuit depth of 41. Despite the moderate circuit complexity, the quantum processor successfully executed ~$10^8$ shots, demonstrating the hardware's capability for large-scale sampling. After symmetry filtering, we obtained approximately 34,500 unique determinants satisfying electron count and spin constraints—representing just 0.003% of the total configurational space. These measurements therefore define a highly concentrated, chemically relevant subspace. H-Couple[39] can be used to add determinants directly connected to this subspace via the electronic Hamiltonian. Alternatively, S-CORE[40] can also be adopted as an alternative determinant recovery strategy.

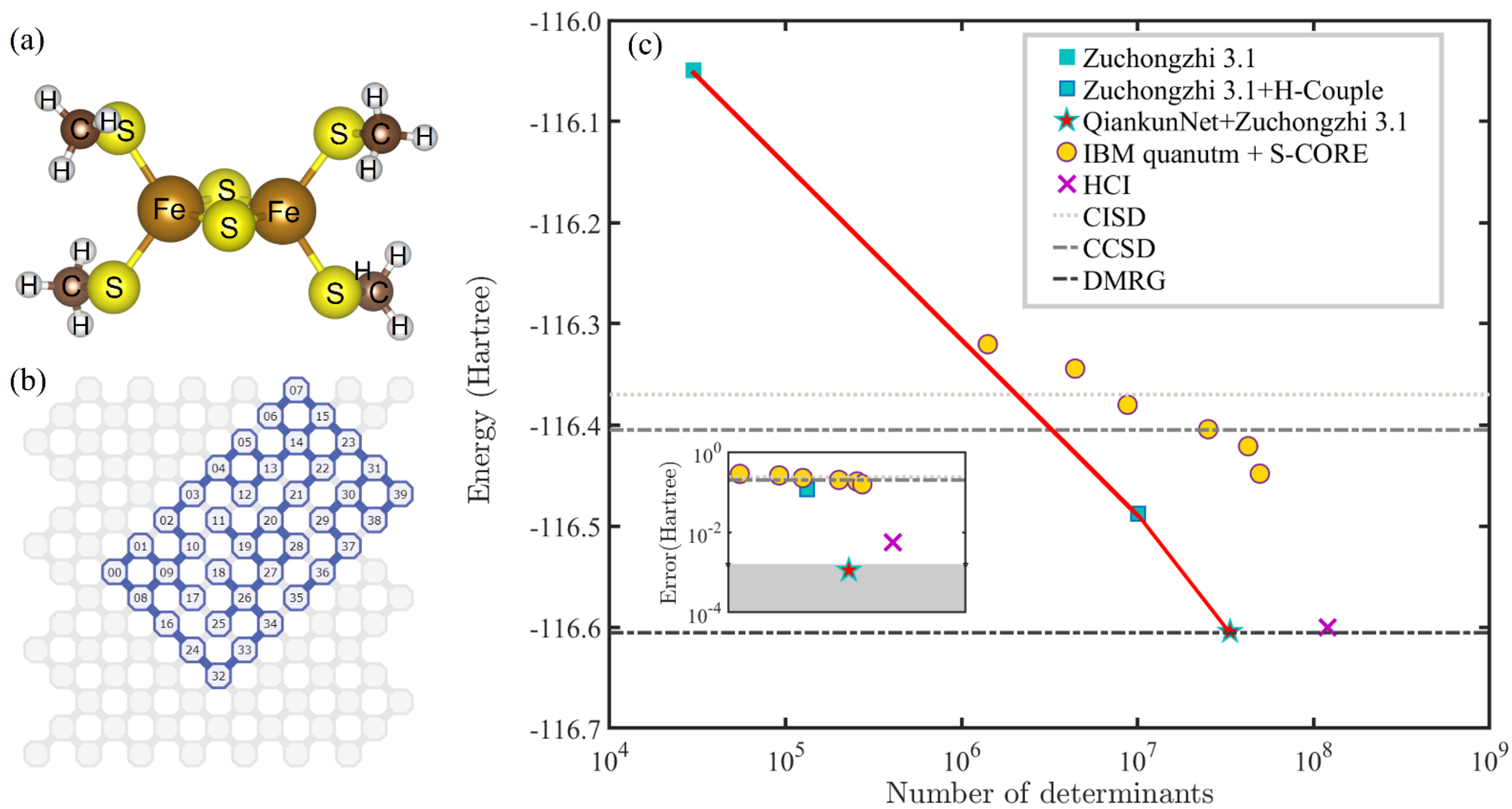

**Fig. 3 Chemical-accuracy solution of the [2Fe-2S] cluster.** (a) Molecular structure of the $[Fe_2S_2(SCH_3)_4]^{2-}$ cluster. (b) Qubit layout on *Zuchongzhi* 3.1 for the 40-qubit experiment. (c) Energy and absolute energy error (inset) versus the number of retained configurations. The number of determinants sampled for local energy calculation is used for QiankunNet. Hardware data for USCI-based sampling on *Zuchongzhi* 3.1 and LUCJ-based sampling on IBM hardware are shown together with classical baselines. The shaded region marks chemical accuracy ($1.6\times10^{-3}$ Ha). Quantum results: QiankunNet-QSCI (QiankunNet+*Zuchongzhi* 3.1: red star), *Zuchongzhi* 3.1 with USCI ansatz (*Zuchongzhi* 3.1: cyan squares) and H-Couple expansions (*Zuchongzhi* 3.1+H-couple: cyan-outlined square), IBM Quantum with LUCJ ansatz with S-CORE (IBM Quantum+S-CORE: yellow circles, magenta outline). Classical benchmarks: heat-bath selected CI (HCI, purple crosses)[40], configuration interaction with singles and doubles (CISD, gray dotted), coupled cluster with singles and doubles (CCSD, gray dashed), and extrapolated DMRG (black dash-dot)[41]. Inset, expanded view near the chemical-accuracy threshold.

As shown in Fig. 3, QiankunNet-QSCI attains ground-state energies within the chemical-accuracy threshold ($1.12 \pm 0.30 \times10^{-3}$ Ha) compared to the DMRG reference energy with a converged bond dimension extrapolation[41]. It significantly surpasses classical algorithms such as CCSD (dashed) and CISD (dotted). Therefore, USCI curbs shot waste by concentrating probability on chemically dominant determinants, while QiankunNet breaks QSCI's truncation ceiling by adaptively boosting under-represented yet important

configurations—yielding accurate, resource-aware performance on a chemically relevant 40-qubit problem.

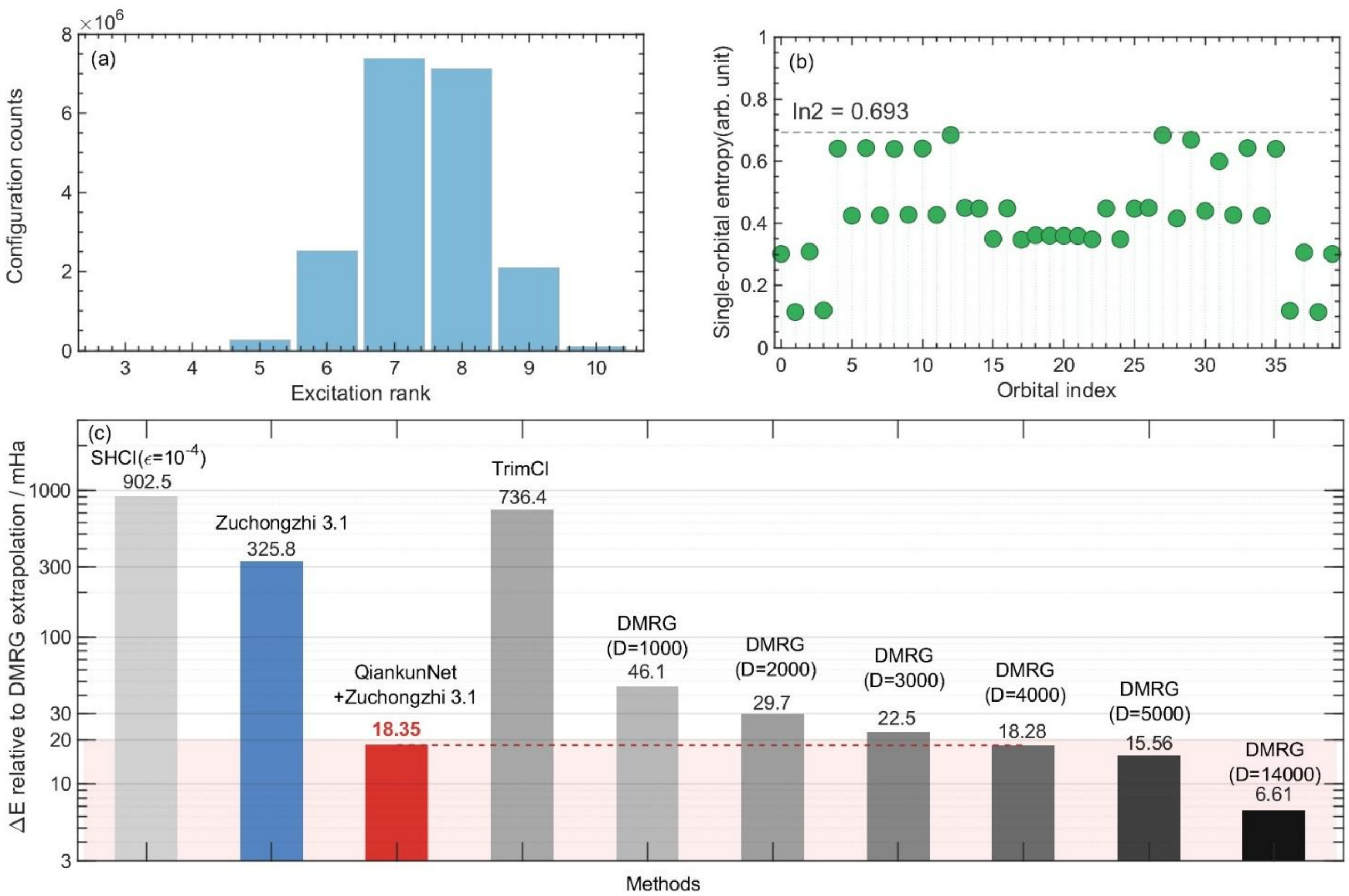

**Fig. 4 Wavefunction analysis of the [2Fe-2S] ground state and benchmark comparison for the P-cluster.** (a) Distribution of retained configurations by excitation rank relative to the Hartree-Fock reference. (b) Single-orbital entropies for the 40 spin orbitals. (c) Energy errors for the P-cluster CAS(114e,73o) model, referenced to DMRG extrapolation,[31] using different models including SHCI ($\epsilon$=$10^{-4}$), TrimCI[42], DMRG (D=1000, 2000, 3000, and 4000)[30], DMRG (D=5000)[43] and DMRG (D=14000)[31].

With a chemically accurate wavefunction in hand, we can study the electron correlation and orbital entanglement properties. Fig. 4(a) shows the distribution of electronic configurations by excitation level relative to the Hartree-Fock reference. The wavefunction is dominated by configurations with 7-8 electronic excitations, demonstrating the inherently multi-configurational nature of [2Fe-2S]. The minimal contributions from both lower and higher excitation levels underscore that intermediate excitations are essential for capturing the system's electronic structure—a hallmark of strong correlation that single-reference methods cannot adequately describe. Fig. 4(b) presents single-orbital entropies for all 40 molecular spin orbitals, quantifying the uncertainty in orbital occupancy and thus their contribution to electron correlation. Several orbitals approach the theoretical maximum entropy of $\ln(2) \approx 0.693$ and others show substantially lower values, revealing a heterogeneous correlation landscape. The entropy profile is consistent with a chemically differentiated active space in which a subset of Fe/S-derived orbitals carries most of the static correlation while other orbitals play a more secondary, dynamical role[28]. We also map the entanglement network through mutual information (Fig. S17). The dominant correlations are not confined to one metal center or to only nearest-neighbor couplings; instead, they are distributed across the $Fe_2$-$S_2$ core and mediated through a small number of highly connected orbitals. This picture is consistent with an electronically delocalized, multi-center description of the cluster.

**Extension to 146-Qubit P-Cluster of Nitrogenase.** The P-cluster ($P^N$ state, [8Fe7S]) is the electron-transfer mediator in biological nitrogen fixation catalyzed by nitrogenase and represents one of the most challenging targets in quantum chemistry due to its eight iron centers and complex spin-coupling topology[30]. We describe the P-cluster electronic structure using a CAS(114e, 73o) active space, yielding 146 spin-orbitals—among the largest metalloenzyme active spaces treated on quantum hardware to date. Under uniform sampling—a useful proxy for the noise-dominated limit where the hardware output approaches a nearly flat bitstring distribution—the probability of finding a configuration with the correct particle number is $P_u = \binom{73}{57}^2 / 2^{146} \approx 3 \times 10^{-13}$, corresponding to far below one valid sample in $10^8$ shots. As a very rough estimation, to at least beat the uniform random sampling, the number of CNOT gates should be smaller than $N_g$, which satisfies $P_q = F_q^{N_g} > P_u$. Here, the average two-qubit fidelity $F_q \approx 0.992$ at the operating condition of *Zuchongzhi* 3.1. Therefore, $N_g$ should not be larger than about 3600, otherwise all signals are expected to be buried in noises. To achieve a useful quantum sampling, further reducing the number of two-qubit gates by at least one order of magnitude is desirable.

To make the quantum circuit as compact as possible, we exploit spin factorization: the 146 spin-orbitals are partitioned into α and β channels, each encoded on a separate 73-qubit register of *Zuchongzhi* 3.1. Entanglement between two spin channels will be recovered via H-Couple and QiankunNet refinement. USCI circuits are built based on 51 determinants prescreened from SHCI, which leads to 37 active excitations in α-spin versus 31 in β-spin. Such a protocol leads to a α-spin circuit with 380 single-qubit gates and 154 CZ gates in 12 layers and a β-spin circuit with 693 single-qubit gates and 146 CZ gates in 6 layers. Each circuit is executed for $10^8$ measurement shots, yielding 1345 (α) and 3335 (β) unique particle-number-conserving configurations. Compared to only $3\times10^{-5}$ valid particle-conserving strings from $10^8$ shots in uniform sampling, this is a significant improvement. More importantly, *QiankunNet* converts these compact, spin-separated quantum samples into a quantitatively accurate many-electron wavefunction. For this system, we don't have a converged FCI result. Even the highest-accuracy DMRG calculation with a bond dimension of 14000 using 48 GPU cards give an energy more than 6 mHa above the extrapolated value[31]. As shown in Fig. 4c, if we use the DMRG extrapolated energy as a reference, the error of QSCI based on *Zuchongzhi* 3.1 is already lower than those of classical methods such as TrimCI.[42] After QiankunNet refinement, the accuracy is further increased to be close to DMRG(D=4000)[30] and DMRG(D=5000).[43]

In summary, by combining a compact ansatz specially designed for quantum sampling together with the QiankunNet-QSCI workflow, accurately solving the electronic structure problem for strongly correlated systems with up to 146 spin-orbitals becomes possible. These results demonstrate that near-term quantum hardware can be useful for practical chemical problems. As hardware progresses, this framework will absorb advances in gate fidelity, coherence, and connectivity, which naturally leads to extensions to excited states, reaction pathways, and response properties—opening opportunities in catalysis, drug discovery, and materials design. Compared to classical methods, such as DMRG, QiankunNet-QSCI is expected to have a better scalability, since the number of qubits required is equal to the number of active orbitals. Therefore, QiankunNet-QSCI shifts the discussion of quantum computing from distant prospects to present capability—delivering scientific value today while establishing architectural patterns for near future's quantum-accelerated discovery.

**Acknowledgements**
This work is supported by the National Natural Science Foundation of China (22393913, 92570112, 22303090), the Strategic Priority Research Program (XDB0450101) and the robotic AI-Scientist platform of the Chinese Academy of Sciences, and the USTC Supercomputing Center.

# Materials and Methods

## 1. Unitary Selected Configuration Interaction Ansatz

The Unitary Selected Configuration Interaction (USCI) ansatz is designed to focus measurement probability on the chemically dominant determinant sector, rather than to approximate the full wavefunction uniformly across Hilbert space. In this sense, USCI should be viewed as the front-end sampler of the QiankunNet-QSCI workflow: its role is to generate a compact, physically meaningful set of determinants that can subsequently be refined by subspace diagonalization and neural-network completion.

In its general form, the USCI wavefunction $|\Psi_{\mathrm{USCI}}(\theta,\kappa)\rangle$ is written as:

$$|\Psi_{\mathrm{USCI}}\rangle = e^{\hat{\kappa}} e^{\hat{T}_s} |\Phi_0\rangle \tag{1}$$

Where $|\Phi_0\rangle$ is typically the Hartree-Fock determinant. The operator $\hat{\kappa}$ performs orbital relaxation,

$$\hat{\kappa} = \sum_{p>q} \kappa_{pq}(a_p^\dagger a_q - a_q^\dagger a_p) \tag{2}$$

In some cases, this operator is not implemented to further reduce the circuit depth. $\hat{T}_s$ introduces correlation by encoding a selected set of excitation patterns defined with respect to the reference determinant,

$$\hat{T}_s = \sum_{K\in\mathcal{S}} \theta_K\left(\tau_K - \tau_K^\dagger\right) \tag{3}$$

Here, $\mathcal{S}$ denotes a compact determinant set obtained from a low-cost classical prescreening step, and $\tau_K$ denotes the excitation operator that connects $|\Phi_0\rangle$ to determinant $|D_K\rangle$. Instead of directly implementing high order excitations, we decompose the occupation changes from $|\Phi_0\rangle$ to $|D_K\rangle$ into symmetry-preserving single- and double-excitation generators and compile them into reusable circuit blocks subject to hardware-connectivity and depth constraints. Expressivity can be enhanced by stacking multiple parametrized blocks.

Parameter optimization proceeds through a hybrid quantum-classical loop. For each parameter set, repeated measurements of the USCI state produce a focused distribution over determinants, which are symmetry filtered and used to define a projected Hamiltonian subspace. This subspace is diagonalized with a Davidson solver, and the lowest eigenvalue is fed back to a derivative-free optimizer, such as COBYLA, to update the circuit parameters until convergence. Classical USCI simulations are performed with our $Q^2$Chemistry platform[44]. HF, CISD, CCSD, and FCI reference calculations are carried out with PySCF[45], and HCI calculations with Dice[46]. Source code and simulation data are provided in the associated repository[47].

## 2. Quantum hardware implementation and experimental parameters

We implemented the LUCJ and USCI ansatz on the *Zuchongzhi* 3.1 superconducting quantum processor. The purpose of the hardware stage is not to solve the full electronic problem by direct variational optimization alone, but to generate an informative determinant pool for subsequent QiankunNet refinement and, when used, Hamiltonian-coupled expansion. For the $H_{10}$ benchmark, both ansatzes target the same (10e,10o) active-space problem. However, the compiled supports differ: the LUCJ circuit acts on the full 20-qubit register, whereas the USCI circuit acts non-trivially only on an 8-qubit support because the dominant selected excitations for this benchmark are confined to 8 spin orbitals. Table 1 therefore distinguishes the full electronic problem size from the compiled active support.

For the [2Fe–2S] cluster, the USCI circuit contains primarily single excitations together with a limited set of double excitations, for a total of 49 variational parameters. The logical ansatz-level circuit has depth 41, while the compiled native-gate circuit executed on *Zuchongzhi* 3.1 has depth 137, with 795 single-qubit gates and 193 CZ gates. We used 40 qubits and accumulated approximately $10^8$ measurement shots for determinant sampling. This distinction between logical ansatz depth and compiled native-gate depth is important and is kept explicit in what follows.

For the P-cluster, we employed a spin-factorized USCI construction in the CAS(114e,73o) active space, treating the α and β-spin sectors separately. Each spin channel is represented by a 73-qubit USCI circuit built from the leading 51 configurations of a preliminary SHCI reference wavefunction. The compiled α-spin circuit contains 380 single-qubit gates and 154 CZ gates arranged in 12 CZ layers, whereas the compiled β-spin circuit contains 693 single-qubit gates and 146 CZ gates in 6 CZ layers. Each circuit was executed for $10^8$ shots. The α circuit yielded 1,345 particle-number-conserving configurations averaged over 30 experimental repetitions, and the β circuit yielded 3,335 such configurations averaged over 60 repetitions. These two configuration pools are subsequently combined in the $\alpha \otimes \beta$ determinant space before QiankunNet training and subspace diagonalization.

**Table 1 Hardware metrics for the $H_{10}$ benchmark, distinguishing full problem size from compiled active support.** $H_{10}$ is treated as a 20-qubit (10e,10o) active-space problem for both ansatzes. For USCI, however, the compiled circuit acts non-trivially on only 8 qubits because the dominant selected excitations are confined to an 8-spin-orbital support in this benchmark. The table reports executed hardware metrics for the compiled circuits.

| Metric | LUCJ | USCI |
|---|---|---|
| Electronic problem size(qubits) | 20 | 20 |
| Compiled active support(qubits) | 20 | 8 |
| Number of Couplers | 27 | 10 |
| Average single-qubit gate error (SQ-XEB) | 0.127% | 0.109% |
| Average CZ gate error (CZ-XEB) | 0.997% | 0.534% |
| Read-out error | 0.735% | 0.552% |
| Total single-qubit gates in circuit | 1192 | 503 |
| Total CZ gates in circuit | 330 | 142 |
| Circuit depth | 255 | 134 |
| Experimental shots | 1e6 | 1e6 |

**Table 2 Hardware metrics for the [2Fe–2S] benchmark on *Zuchongzhi* 3.1.** The table reports compiled native-gate metrics for the executed USCI circuit, including qubit count, coupler count, gate errors, total native gate numbers, compiled depth, and total measurement shots. These values describe the circuit after hardware compilation and should therefore be distinguished from the shallower logical ansatz representation discussed in the main text.

| Metric | USCI |
|---|---|
| Number of qubits | 40 |
| Number of Couplers | 46 |
| Average single-qubit gate error (SQ-XEB) | 0.117% |
| Average CZ gate error (CZ-XEB) | 0.758% |
| Read-out error | 0.675% |
| Total single-qubit gates in circuit | 795 |
| Total CZ gates in circuit | 193 |
| Logical ansatz-level depth | 41 |
| Compiled native-gate depth | 137 |

1e8

Experimental shots

**Table 3 Hardware metrics for the spin-factorized P-cluster benchmark on *Zuchongzhi* 3.1.** Separate compiled metrics are reported for the α- and β-spin USCI circuits, including pre-compilation logical gate counts, compiled native-gate counts, CZ layer numbers, particle-number-conserving sampled configurations, and total shots. The two spin sectors are executed independently and later recombined during classical post-processing.

| Metric | USCI (Alpha) | USCI (Beta) |
|---|---|---|
| Number of qubits | 73 | 73 |
| Pre-compilation 1q gates | 211 | 211 |
| Pre-compilation 2q gates | 308 | 300 |
| Pre-compilation circuit depth | 37 | 35 |
| Compiled single-qubit gates | 380 | 693 |
| Compiled CZ gates | 154 | 146 |
| CZ gate layers | 12 | 6 |
| Particle-conserving configs | 1,345 (30×) | 3,335 (60×) |
| Experimental shots | $10^8$ | $10^8$ |

## Supplementary Materials

# Transformer-refined quantum sampling for strongly correlated electronic structure

# Content:

# 1. Unitary Selected Configuration Interaction (USCI) Ansatz

## 1.1 Wavefunction Formulation

In the USCI formulation, the wavefunction is generated by applying the unitary operators to a reference determinant, typically the Hartree-Fock state:

$$|\Psi\rangle = e^{-\hat{k}} e^{-\hat{R}} |\Phi_0\rangle \tag{1}$$

This form is closely related to the exponential parametrization used in multiconfigurational self-consistent field (MCSCF) theory[1]. The ansatz contains two conceptually distinct ingredients $\hat{k}$ and $\hat{R}$. The first is an orbital-rotation operator $\hat{k}$, which mixes molecular orbitals through an anti-Hermitian generator and plays a role analogous to orbital optimization in conventional multiconfigurational methods. The second is a selected-excitation operator $\hat{R}$, which introduces correlation by coupling the reference determinant to a compact set of important excited determinants. This separation is useful both conceptually and practically: orbital rotations define a flexible one-particle basis, whereas selected excitations concentrate variational freedom on the physically relevant sector of configuration space.

$$\hat{k} = \sum_{p>q,\sigma} k_{pq} \left( \hat{a}^{\dagger}_{p\sigma} \hat{a}_{q\sigma} - \hat{a}^{\dagger}_{q\sigma} \hat{a}_{p\sigma} \right) \tag{2}$$

This term introduces single-particle excitations to the wavefunction. The state-transfer operators $\hat{\mathrm{R}}$ are defined as:

$$\hat{R} = \sum_{K>J, J\leq n} R_{KJ} \left( |K\rangle\langle J| - |J\rangle\langle K| \right) \tag{3}$$

The state-transfer operators are built by comparing the occupation patterns of the reference determinant and an excited determinant and identifying the creation and annihilation operators required to transform one into the other. In this way, the fermionic content of each excitation is made explicit before qubit encoding. Although the illustrative expressions below focus on the double-excitation case for clarity, the formal construction is not restricted to doubles: single, double, triple, and higher excitation orders are all admissible in principle, given that they satisfy the relevant symmetry constraints.

## 1.2 Unitary factorization and determinant-pair rotations

To obtain a circuit-compatible representation, the selected-configuration unitary is first expressed as a product of elementary rotations acting within selected determinant pairs. In the determinant basis, the anti-Hermitian generator can be written as:

$$\hat{R} = \sum_{(K,J)\in\mathcal{S}} R_{KJ} (|K\rangle\langle J| - |J\rangle\langle K|), \quad R_{KJ} \in \mathbb{R} \tag{4}$$

Where $\mathcal{S}$ denotes the selected set of determinant pairs. The corresponding unitary transformation is

$$e^{-\hat{R}} = \exp\left[ -\sum_{(K,J)\in\mathcal{S}} R_{KJ} (|K\rangle\langle J| - |J\rangle\langle K|) \right] \tag{5}$$

Equivalently, after a first-order Trotter factorization, this unitary can be approximated as an ordered product of local determinant-pair rotations,

$$e^{-\hat{R}} \approx \mathcal{T} \prod_{(K,J)\in\mathcal{S}} \exp\left[ -R_{KJ} (|K\rangle\langle J| - |J\rangle\langle K|) \right] \tag{6}$$

where $\mathcal{T}$ denotes the prescribed operator ordering. Each elementary factor performs a Givens-like amplitude rotation in the two-dimensional subspace spanned by $|J\rangle$ and $|K\rangle$. Suppose, for example, that two determinants $|J\rangle$ and $|K\rangle$ differ by a double excitation, in which electrons are removed from occupied spin orbitals $t$, $r$ and created in spin orbitals $p$, $q$. We define the corresponding fermionic excitation operator as

$$\hat{\tau}_{KJ} = \eta_{KJ}\hat{a}_p^\dagger \hat{a}_q^\dagger \hat{a}_r \hat{a}_t, \quad \hat{\tau}_{KJ}|J\rangle = |K\rangle \tag{7}$$

where $\eta_{KJ} = \pm 1$ is the fermionic phase determined by the ordering convention of Slater determinants. With the determinant projector is $\hat{P}_J = |J\rangle\langle J|$, the exact transition operator between the two determinants is $|K\rangle\langle J| = \hat{\tau}_{KJ}\hat{P}_J$. The associated exact two-configuration unitary is therefore:

$$\hat{\mathrm{U}}_{\mathrm{KJ}} = \exp\left[\theta_{KJ}(|K\rangle\langle J| - |J\rangle\langle K|)\right] = \exp\left[\theta_{KJ}\left(\hat{\tau}_{KJ}\hat{P}_J - \hat{P}_J\hat{\tau}_{KJ}^\dagger\right)\right] \tag{8}$$

Here $\hat{\tau}_{KJ}^\dagger = \eta_{KJ}\hat{a}_t^\dagger \hat{a}_r^\dagger \hat{a}_q \hat{a}_p$, up to the same determinant-ordering convention. The generator in Eq. (8) is anti-Hermitian; hence its exponential is unitary for real $\theta_{KJ}$. This operator does not merely introduce a phase. Instead, it mixes the amplitudes of $|J\rangle$ and $|K\rangle$ while leaving all determinants outside this two-dimensional subspace unchanged.

Sequential application of these exact determinant-pair rotations gives:

$$\hat{\mathrm{U}}_{\mathrm{USCI}} = \prod_{(K,J)\in\mathcal{S}} \exp\left[-R_{KJ}(|K\rangle\langle J| - |J\rangle\langle K|)\right] = \prod_{(K,J)\in\mathcal{S}} \exp\left[\theta_{KJ}\left(\hat{\tau}_{KJ}\hat{P}_J - \hat{P}_J\hat{\tau}_{KJ}^\dagger\right)\right] \tag{9}$$

Equation (9) represents the formally exact selected-CI rotation in the determinant basis. However, the explicit projectors $\hat{P}_J$ make the operator highly nonlocal after fermion-to-qubit mapping and would lead to expensive multi-controlled many-body operations. In the practical USCI ansatz, we therefore adopt an ansatz-level approximation: the determinant projectors are omitted, and only the corresponding anti-Hermitian excitation generators are retained. This converts the exact determinant-pair rotation into a compact UCC-like excitation unitary:

$$\hat{\tau}_{KJ}\hat{P}_J - \hat{P}_J\hat{\tau}_{KJ}^\dagger \longrightarrow \hat{\tau}_{KJ} - \hat{\tau}_{KJ}^\dagger \tag{10}$$

This approximation preserves particle number and the selected excitation structure, while substantially reducing circuit depth. The resulting excitation unitary is no longer restricted to a single determinant pair; rather, it variationally mixes all configurations connected by the same excitation pattern. The associated leakage outside the originally selected two-determinant subspace is controlled by determinant prescreening, excitation truncation, circuit-depth constraints, and variational parameter optimization.

Starting from the reference determinant $|\Phi_0\rangle$, the practical USCI wavefunction is therefore defined as

$$|\Psi_{\mathrm{USCI}}\rangle = e^{-\hat{k}} \prod_{(K,J)\in\mathcal{S}} \exp\left[\theta_{KJ}\left(\hat{\tau}_{KJ} - \hat{\tau}_{KJ}^\dagger\right)\right] |\Phi_0\rangle \tag{11}$$

where $e^{-\hat{k}}$ denotes the orbital-rotation unitary, and $\theta_{KJ}$ are variational parameters associated with the selected excitation channels. The double excitation shown above is only an illustrative example; in actual USCI constructions, the selected channels may include symmetry-allowed single, double, or decomposed higher-order excitations, depending on the determinant prescreening and hardware constraints.

### 1.3 **Quantum Circuit Design and Configuration Selection**

After the fermionic operators have been defined, they are mapped to qubit operators through the Jordan-Wigner transformation[2]:

$$\hat{a}_p \longrightarrow \frac{1}{2}\left(X_p + iY_p\right)\prod_{i=0}^{p-1} Z_i \ , \hat{a}_p^\dagger \longrightarrow \frac{1}{2}\left(X_p - iY_p\right)\prod_{i=0}^{p-1} Z_i \tag{12}$$

This mapping provides an explicit qubit representation of the selected orbital-rotation and state-transfer operators, thereby connecting the determinant-based construction of USCI to an executable quantum circuit.

In practice, we employ two approximations. First, we perform a low-cost classical prescreening step, based on CISD or a loosely converged HCI/SHCI calculation[3], and retain only the most significant determinants ranked by amplitude. Excitation operators are then generated only among these selected configurations rather than over the full CI manifold. Second, higher-order excitations are factorized into combinations dominated by single and double excitations. For example, a triple excitation is approximated by one double plus one single, a quadruple by two doubles, and a quintuple by two doubles plus one single. These reductions substantially decrease circuit depth and two-qubit gate counts while preserving the dominant low-rank correlation structure.

$$|\Psi\rangle = \prod_{l=1}^{L} e^{-\hat{\kappa}_l(\Theta)} e^{-\hat{R}_l(\Theta)} |\Phi_0\rangle \tag{13}$$

The resulting USCI workflow should therefore be understood as a deliberately biased but chemically motivated compression strategy. It concentrates variational resources on the determinant sector most relevant to the target state and thereby produces a circuit that is both expressive and compatible with present-day superconducting hardware.

The construction of the USCI circuit begins from an inexpensive classical calculation, typically CISD or a loosely converged semi-stochastic HCI run, which provides an approximate wavefunction and a ranked list of determinants. All determinants are sorted according to the absolute value of their coefficients, and the dominant determinant is chosen as the reference state. An amplitude cut-off is then imposed, and only determinants above this threshold are retained. The excitation operators connecting these retained determinants to the reference form the elementary building blocks of the circuit.

Although the optimal cut-off is system dependent, classical benchmarks across a range of strongly correlated molecules indicate that a relatively small set of leading determinants is often sufficient to generate shallow circuits with competitive variational quality. In systems with near-degenerate orbitals, groups of determinants may carry comparable weights; such quasi-degenerate determinants are deliberately preserved so that the principal correlation channels remain represented already at the circuit-construction stage. At the same time, the final time-evolved wavefunction is not confined to the initial seed set itself. Through the action of the selected excitation operators and the resulting qubit dynamics, the ansatz can explore a significantly larger correlated neighborhood while remaining shallow enough for NISQ execution.

### 1.4 **Circuit Implementation and Parameter optimization**

In practice, we adopt a generalized single-and double-excitation architecture[4], which offers a useful balance between depth and the ability to capture the dominant correlation structure. Because the pool of candidate excitations is large, a qubit could in principle become coupled to many others, thereby inflating the two-qubit gate count and introducing large routing overhead after compilation.

This issue is especially important on superconducting hardware such as *Zuchongzhi* 3.1, whose native connectivity is restricted by a square-lattice topology[5]. To match the ansatz to this hardware constraint, we build the circuit incrementally while maintaining a connectivity table. Each time a single or double excitation is inserted, the table is updated; once a qubit has reached the allowed number of interaction partners, no further cross-qubit excitations involving that qubit are introduced within the same block. The expressive power sacrificed by this degree-capping strategy is then partially recovered by stacking multiple structurally identical entangling blocks with independent parameters. In this way, the compiled circuit preserves the physically motivated sparsity of the ansatz while remaining compatible with the limited connectivity of the device.

The final circuit can therefore be written as a sequence of parameterized gates $G(\theta)$:

$$G(\theta) = G(\theta_1)G(\theta_2) \ldots G(\theta_n) \tag{14}$$

where each gate represents one local unitary transformation derived from the selected excitation structure. The gates are applied sequentially to implement the orbital-rotation and determinant-transfer operations defining the USCI ansatz.

Parameter optimization is carried out through a hybrid quantum-classical loop. For a given set of variational parameters, the USCI circuit is executed on a quantum processor or a high-fidelity simulator, and repeated projective measurements produce a probability distribution over sampled Slater determinants. These bit strings are then post-processed to recover symmetry-consistent configurations and low-rank excitations, yielding an expanded determinant list that captures the most relevant sector of Hilbert space. From this determinant set we assemble the projected Hamiltonian and overlap matrices and solve the resulting generalized eigenvalue problem with an in-house Davidson diagonalization routine optimized for large sparse subspaces. The lowest eigenvalue serves as the objective for a derivative-free classical optimizer, and in practice we use COBYLA to avoid costly gradient evaluations on noisy hardware.

The optimizer then proposes a new parameter set, the circuit is re-executed, and the cycle is repeated until the energy change falls below the chosen convergence threshold. This closed-loop procedure makes the role of the quantum processor precise: it supplies a focused determinant distribution, while the classical stage converts that distribution into an improved variational estimate of the target state. The purpose of USCI within QiankunNet-QSCI is therefore specific but essential: it is not introduced as a stand-alone route to exact electronic structure, but as a compact hardware-level mechanism for enriching the determinant sector passed to the downstream classical refinement.

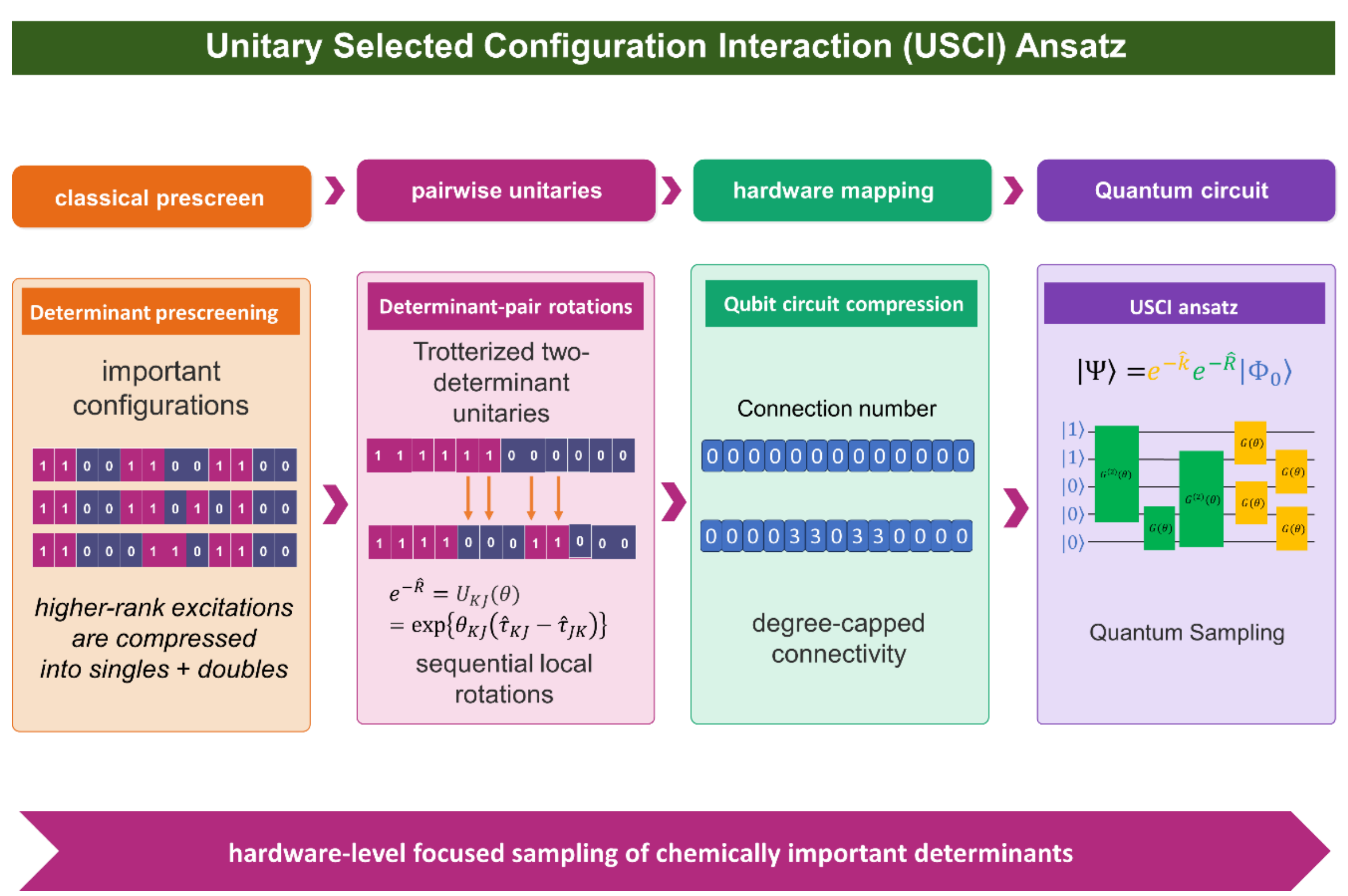

**Fig.S1 Workflow for constructing the USCI ansatz.** Chemically important determinants are first selected by classical prescreening, then encoded as sequential determinant-pair unitaries, compressed into a hardware-compatible qubit circuit, and compiled into the final USCI ansatz for focused quantum sampling.

## 2. USCI versus LUCJ: Theory, Circuit Construction, and Benchmark Results

To place USCI in context, we compare it with the Local Unitary Cluster-Jastrow (LUCJ) ansatz, which has emerged as an important hardware-aware variational form for near-term quantum chemistry. The purpose of this comparison is not to claim universal dominance of one ansatz over the other, but to clarify the regime targeted by USCI and to explain why determinant-focused circuit construction can be advantageous for the hybrid workflow developed in this study. In particular, LUCJ is designed as a flexible hardware-conscious variational ansatz with layered orbital-rotation and correlator blocks, whereas USCI is constructed around a prescreened determinant manifold and therefore emphasizes compactness and chemical selectivity at the sampling stage.

### 2.1 Analytical comparison of unitary generators and gate counts

One layer of the LUCJ ansatz prepares[6]

$$\left|\Psi_{LUCJ}\right\rangle = e^{\hat{K}} e^{i\hat{J}} e^{-\hat{K}} |\Phi_0\rangle \tag{15}$$

where the first unitary $\hat{K} = \sum_{pq,\sigma} K_{pq}\, \hat{a}^{\dagger}_{pq} \hat{a}_{q\sigma}$ performs an orbital rotation and the second $\hat{J} = \sum_{pq,\sigma\sigma'} J_{pq;\sigma\sigma'}\ \hat{n}_{p\sigma}\hat{n}_{q\sigma}$ is a number-number(Jastrow) correlator. Stacking $L$ identical layers systematically increases the expressive power of the ansatz and, in principle, approaches the exact limit as $L \to \infty$ limit. In practice, the associated parameters may be initialized from classical amplitudes, such as those obtained from CCSD.[5]

In LUCJ, the orbital-rotation block typically requires a mesh of Givens-type transformations, and the resulting gate count grows with the size of the orbital space even before hardware routing is considered. In USCI, the effective gate count depends primarily on the number of retained determinant couplings rather than on the full combinatorial size of the active space. When the selected determinant manifold remains compact, the corresponding circuit can therefore be substantially shallower. This feature is particularly important here, because the goal is not merely to lower a standalone variational energy, but to generate a focused determinant distribution that will subsequently seed QiankunNet refinement.

**Table S1: Qualitative comparison of parameter count and circuit-depth scaling for LUCJ and USCI.** LUCJ distributes variational freedom over orbital-rotation and correlator blocks whose cost scales with the size of the active space and the number of layers. USCI combines orbital rotations with selected determinant-transfer operations built from a compact prescreened manifold of size mmm, so its cost scales with the retained determinant structure rather than with the full CI space. When mmm is small, this leads to substantially lower circuit depth and fewer entangling operations, especially on hardware with restricted connectivity.

| Quantity | LUCJ (per layer) | USCI (per layer) |
|---|---|---|
| Variational parameters | $O(N^2)$ from $K_{pq}$ plus $O(N^2)$ from $J_{pq}$ | $O(m)$ from $\hat{R}_l(\Theta)$ plus $O(N)$ from $\hat{k}_l(\Theta)$ where m is the number of selected configurations |
| Two-qubit gates | All-to-all: $2N(2N-1)/2$<br>Local form: linear in $N$ | $O(m+N)$ |
| Circuit depth | Orbital-rotation block: 1+N<br>Jastrow block: constant=4 on | Orbital-rotation block: 4<br>State-transfer block: $O(m)$ |

square/hex/linear topologies,8 on
heavy-hex

For LUCJ, the number of variational parameters is determined by the orbital-rotation and correlator tensors within each layer, and the gate structure remains comparatively dense even in local forms designed to reduce SWAP overhead. For USCI, the number of parameters is set by the orbital-rotation terms together with the selected determinant couplings, so the effective scaling depends on the size of the retained determinant set mmm, rather than on the full CI manifold. Accordingly, when mmm remains modest, USCI provides a more economical route to a hardware-compatible circuit.

### 2.2 Numerical performance on representative molecules

To assess whether USCI improves not only circuit compactness but also the quality of the determinant pool passed to downstream classical refinement, we benchmark it against the single-layer LUCJ ansatz for the stretched $N_2$ molecule at a bond length of 3.0 Å, corresponding to a 20-qubit active-space problem. This geometry provides a useful test case because static correlation is already pronounced, so the ansatz must balance expressivity, trainability, and hardware cost.

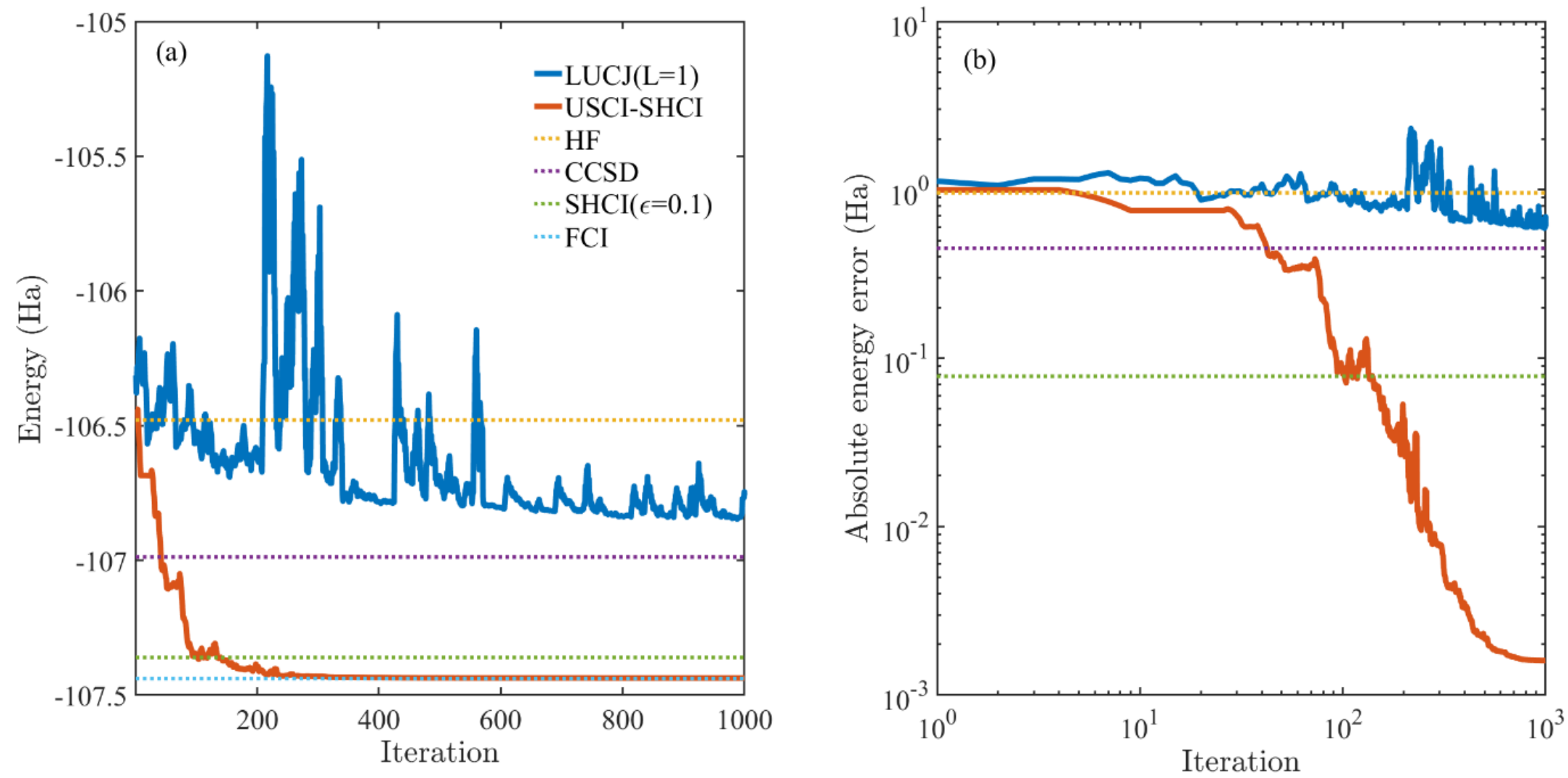

**Fig. S2 Direct comparison of LUCJ and USCI-SHCI for stretched $N_2$ at 3.0 Å.** (a) Ground-state energy versus optimization iteration for L=1 LUCJ and SHCI-seeded USCI. Horizontal lines indicate HF, CCSD, SHCI (ε=0.1), and FCI reference energies. (b) Absolute energy error relative to FCI. USCI-SHCI converges faster and reaches lower error with a substantially more compact circuit than LUCJ.

Fig. S2 shows that USCI achieves a more favorable balance between compactness and convergence than LUCJ. For the SHCI-seeded USCI circuit at L=1, the compiled circuit has depth 115, with 114 single-qubit gates, 142 two-qubit gates, and 38 variational parameters. In contrast, the L=1 LUCJ circuit has depth 135, with 838 single-qubit gates, 246 two-qubit gates, and 117 parameters. The advantage of USCI is therefore not limited to a single metric: it simultaneously reduces circuit depth, gate count, and parameter number. More importantly, the lower-cost USCI circuit also converges more rapidly and more smoothly, indicating that determinant-selected excitations provide a more efficient parametrization of the correlated subspace relevant to this stretched-bond regime.

In our framework, the quantum ansatz is not an end point. Instead, its purpose is to generate a determinant distribution that can be refined further by QiankunNet. Figure S3 therefore addresses the more consequential question for the overall workflow: does a better quantum ansatz only lower the standalone variational energy, or does it also furnish a better initialization for machine-learned wavefunction completion? The answer is the latter. When QiankunNet is initialized with the determinant pool generated from USCI-SHCI, the hybrid optimization reaches chemical accuracy after only 58 iterations. In contrast, initialization from the LUCJ-generated determinant pool requires 436 iterations, and the standalone QiankunNet baseline converges more slowly still. Thus, the advantage of USCI is not merely that it is cheaper to implement; rather, it produces a more informative and chemically focused starting distribution for the subsequent neural refinement.

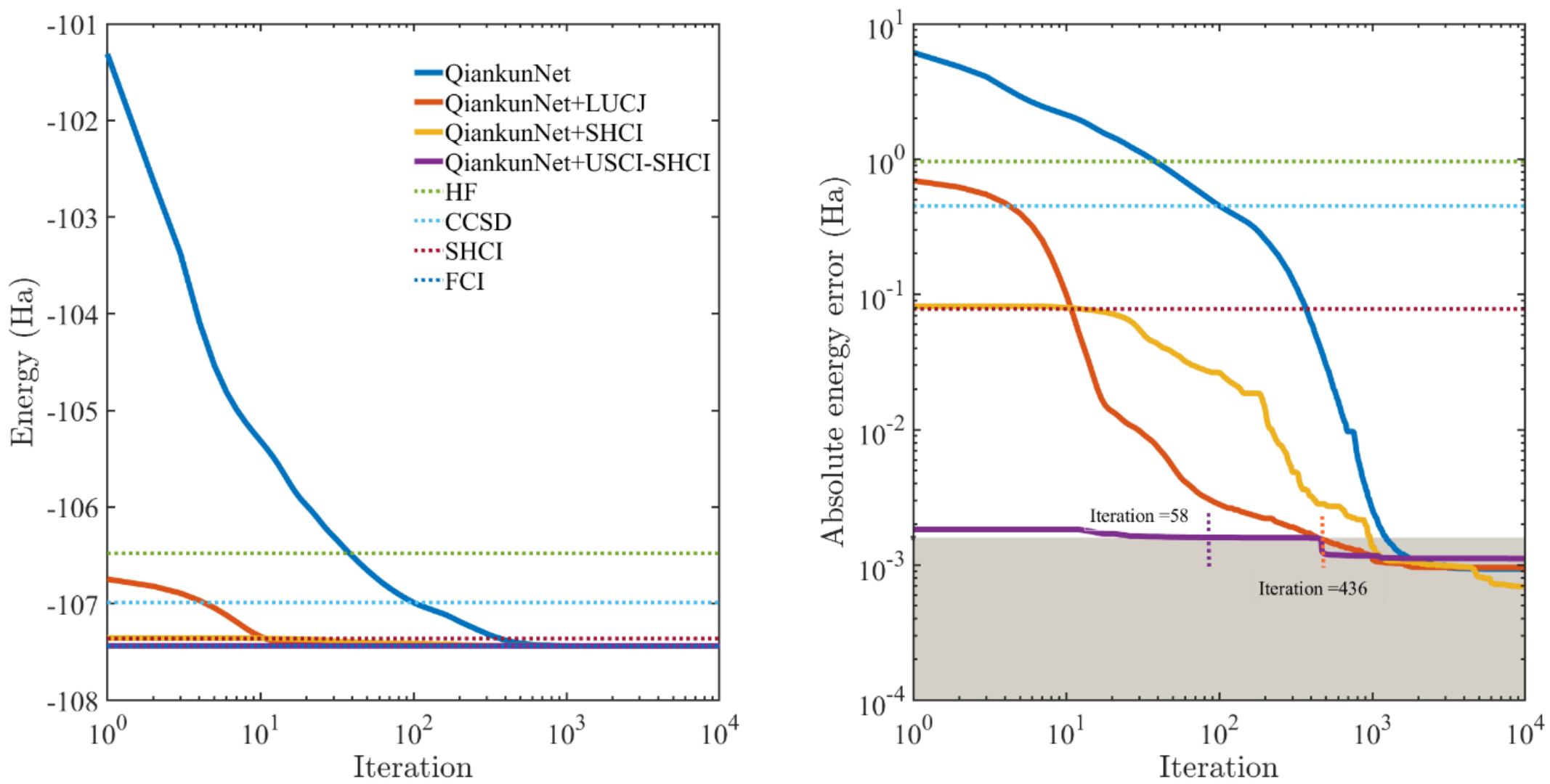

**Fig. S3 Effect of determinant-pool initialization on QiankunNet for stretched $N_2$ at 3.0 Å.** (a) Ground-state energy versus optimization iteration for QiankunNet alone and for hybrid workflows initialized from LUCJ, SHCI, or USCI-SHCI determinant pools. Horizontal lines indicate HF, CCSD, SHCI (ε=0.1), and FCI references. (b) Absolute energy error relative to FCI. The shaded region marks chemical accuracy. USCI-SHCI initialization reaches chemical accuracy after 58 iterations, compared with 436 iterations for LUCJ initialization.

# 3 General performance of determinant-focused sampling

## 3.1 Transferability of USCI across representative strongly correlated molecules

In this section, we examine a chemically diverse benchmark set spanning main-group diatomic, a transition-metal dimer, a π-diradical organic system, and a bioinorganic $[Cu_2O_2]^{2+}$ core. The goal here is not to claim a uniform level of quantitative accuracy across all systems, but to assess whether the same shallow, determinant-selected circuit construction remains transferable across qualitatively different correlation regimes.

**Table S2: Active spaces and benchmark ground-state energies for five representative strongly correlated molecules.** Ground-state energies (Hartree) are reported for CASCI, loose-threshold HCI (ε=0.01 Ha), and USCI with one and two entangling layers. The benchmark set includes $N_2$, $O_2$, $Cr_2$, cyclobutadiene, and $[Cu_2O_2]^{2+}$, spanning closed-shell, open-shell, metal-metal, organic diradical, and metal-ligand correlated regimes. References for active-space definitions and benchmark geometries are given in Refs.[7-12]

| Molecule | Basis Set | Active Space ($N_e$, $N_o$) | Active Orbitals | CASCI (Ha) | HCI (ε=0.01 Ha) | USCI (L=1) (Ha) | USCI (L=2) (Ha) |
|---|---|---|---|---|---|---|---|
| $N_2$ | cc-pVDZ | (10,8) | $3\sigma_g$, $1\pi_u$ (×2), $3\sigma_{g^*}$, $1\pi_{g^*}$ (×2) | -109.034407 | -109.030234 | -109.024859 | -109.025965 |
| $O_2$ ($^3\Sigma_g^-$) | cc-pVDZ | (10,8) | $\pi_g$ (×2), $\pi_{g^*}$ (×2), ($\sigma_u$, $\sigma_{u^*}$) (×2) | -149.700280 | -149.698493 | -149.6933143 | -149.697439 |
| $Cr_2$ | cc-pVDZ-DK | (12,12) | 4s σ, plus five 3d-σ/π/δ bonding & antibonding pairs | -1180.830331 | -1180.287512 | -1180.449050 | -1180.553953 |
| Cyclobutadiene ($C_4H_4$) | cc-pVDZ | (8,8) | Full π manifold: 4 bonding + 4 antibonding π MOs | -153.497381 | -153.460670 | -153.481618 | -153.4907538 |
| $[Cu_2O_2]^{2+}$ | def2-TZVP | (16,12) | Cu $3d^{10}$ + $O_2$ σ, π (2) | -1655.877803 | -1655.284164 | -1655.789736 | -1655.877306 |

Table S2 shows that the same USCI construction can be applied across chemically distinct strongly correlated systems without changing its basic design logic. In every case, the ansatz is built from a compact prescreened determinant set, higher-order excitations are reduced to an effective low-rank representation, and additional flexibility is introduced only by stacking shallow entangling layers.

A single USCI layer already yields non-trivial variational improvements across all five systems, and a second layer systematically lowers the energy further. The magnitude of that improvement is strongly system dependent. For cyclobutadiene and $[Cu_2O_2]^{2+}$, the two-layer ansatz narrows the gap to the CASCI reference substantially; for $Cr_2$, the additional layer also produces a large energy gain relative to the one-layer form, although a sizeable residual gap remains. In contrast, for $N_2$ and $O_2$, the present shallow USCI parametrization improves upon the one-layer result but does not outperform the loose-threshold HCI reference. Accordingly, Table S2 should not be read as evidence that a low-depth USCI circuit uniformly reaches CASCI quality, but rather that the ansatz provides a transferable and systematically improvable low-depth starting point across diverse strongly correlated systems.

To place these results in practical context, Table S3 summarizes the circuit resources required per USCI layer for the same benchmark set.

**Table S3 Per-layer logical resource requirements of the USCI circuit for the five molecular benchmarks.** For each molecule, we report the number of retained configurations after determinant prescreening, the resulting number of variational parameters, and the logical single-qubit gate count, logical two-qubit gate count, and circuit depth for one USCI entangling layer after decomposition into the basic gate set used in our simulations.

| Molecule | USCI circuit parameters per layer | | | | |
|---|---|---|---|---|---|
| | Number of configs | Number of parameters | Number of single qubit gates | Number of two qubit gates | Circuit depth |
| $N_2$ | 9 | 64 | 580 | 604 | 467 |
| $O_2$ ($^3\boldsymbol{\Sigma}_g^-$) | 10 | 74 | 692 | 724 | 680 |
| $Cr_2$ | 13 | 140 | 1208 | 1302 | 1104 |
| Cyclobutadiene $C_4H_4$ | 11 | 93 | 866 | 914 | 803 |
| $[Cu_2O_2]^{2+}$ | 12 | 119 | 1040 | 1114 | 950 |

Table S3 shows that the USCI construction remains compact at the level of determinant selection even as the chemistry changes: only 9–13 configurations are retained in these examples, leading to 64–140 trainable parameters per layer. The resulting logical gate counts and circuit depths are not negligible, but they remain far below what would be expected from an unbiased treatment of the full active-space excitation manifold. Results in Table S3 do not establish that all five systems are immediately suitable for direct hardware execution in their present form; rather, it shows that determinant prescreening and degree-capped circuit construction compress the ansatz into a regime compatible with the hardware-oriented design goals laid out in Section 2.

## 3.2 Sampling bias, noise robustness, and QiankunNet reweighting

We use stretched $N_2$ as a benchmark to examine how finite-depth noise distorts the determinant distribution produced by USCI, how that distortion changes determinant sparsity and energy, and to what extent QiankunNet can exploit even imperfect quantum samples as an initialization for subsequent refinement.

To model the effect of realistic NISQ execution, we consider two sources of measurement bias. The first is gate-induced depolarizing noise, which alters the prepared quantum state before measurement. The second is classical readout error, which perturbs the measured bit strings after projection in the computational basis. The corresponding channels are defined below and are used only as controlled error models for the present $N_2$ benchmark; they are not intended to reproduce the full device-level error profile of a specific processor.

Depolarizing-noise channel: For an n-qubit register let $\mathcal{P}_n = \{I, X, Y, Z\}^{\otimes n}$ be the non-trivial Pauli operators, and the canonical global depolarizing channel with strength $p \in [0,1]$ is the completely-positive, trace-preserving (CPTP) map[13]:

$$D_p^{(n)}(\rho) = (1-p)\rho + \frac{p}{4^n - 1} \sum_{P \in \mathcal{P}_n} P\rho P \tag{16}$$

Read-out error model: Let the ideal measurement project on $\{|0\rangle, |1\rangle\}^{\otimes n}$. Readout noise is classical and fully described by a stochastic matrix:

$$M_{ij} = Pr(out\ i\ |state\ j), \qquad \sum_i M_{ij} = 1 \tag{17}$$

with bit-string $i, j \in \{|0\rangle, |1\rangle\}^{\otimes n}$. Assume each qubit is flipped independently:

$$M = \bigotimes_{q=1}^{n} M^{(q)}, \qquad M^{(q)} = \begin{pmatrix} 1-\varepsilon_0^{(q)} & \varepsilon_1^{(q)} \\ \varepsilon_0^{(q)} & 1-\varepsilon_1^{(q)} \end{pmatrix} \tag{18}$$

were $\varepsilon_0^{(q)}(\varepsilon_1^{(q)})$ is the probability that a prepared $|0\rangle(|0\rangle)$ is mis-reported.

Using this setup, we perform noise-controlled simulations for the 20-qubit $N_2$ active space with 10,000 measurement shots. The FCI result provides the exact reference, while CISD gives a compact but substantially less accurate classical baseline. The noiseless USCI circuit yields a much sparser determinant set than FCI and remains substantially lower in energy than CISD, indicating that the determinant-focused circuit already captures a nontrivial fraction of the correlated manifold using only a small number of retained configurations. More importantly for the present purpose, mild noise has only a limited effect on the USCI output distribution: at p=0.01, the energy changes only slightly relative to the noiseless USCI result, and the determinant count remains essentially unchanged. By contrast, at p=0.02 and especially at p=0.10, both the energy and the retained determinant pool are visibly distorted, indicating that the circuit is no longer sampling only the compact chemically dominant sector.

For sufficiently mild noise the determinant distribution produced by USCI remains structured and sparse enough to be useful as a starting point for further refinement. Once the noise becomes too strong, however, the determinant support broadens, the dominant amplitudes are redistributed, and the sampled subspace loses much of the chemically informative bias that makes the hybrid workflow effective in the first place.

**Table S4 Ground-state energies, determinant counts, and leading configurations for the 20-qubit $N_2$ with bond length 3.0 Å benchmark under controlled depolarizing noise.** Rows compare the FCI reference, the CISD baseline, and USCI simulations with increasing depolarizing-noise strength $p$. For each case we report the total energy, the number of determinants retained above the coefficient threshold used in the analysis, and the five configurations with the largest amplitudes.

| Description | Energy (Ha) | Number of Configs. (\|coeffs\|>1e-15) | Top 5 Configs. | |
|---|---|---|---|---|
| | | | Configs. | Coeffs |
| FCI results | -107.438440 | 7119 | 11111111111001100100 | 0.239134 |
| | | | 11111111110110011000 | 0.239004 |
| | | | 11111111111111000000 | 0.234558 |
| | | | 11111111110011110000 | -0.231115 |
| | | | 11111111111100001100 | -0.231115 |
| CISD results | -107.126850 | 534 | 11111111110110011000 | 0.559455 |
| | | | 11111111110101101000 | 0.517241 |
| | | | 11111111111010010100 | 0.517241 |
| | | | 11111111100111110000 | -0.319336 |
| | | | 11111111111001100100 | 0.302834 |
| USCI simulation without noise | -107.3145479 | 12 | 11111111000011110011 | 0.380622 |
| | | | 11111111110000111100 | -0.374644 |
| | | | 11111111110110110000 | -0.374549 |
| | | | 11111111110101110000 | 0.368757 |
| | | | 11111111001100001111 | -0.335926 |
| USCI simulation with p=0.01 | -107.314501 | 11 | 11111111110000111100 | -0.380641 |
| | | | 11111111001111000011 | 0.374682 |
| | | | 11111111111100001100 | 0.374639 |
| | | | 11111111001100001111 | -0.368774 |
| | | | 11111111000000111111 | 0.335903 |
| USCI simulation with p=0.02 | -106.629069 | 16 | 11111111101100100111 | -0.502496 |
| | | | 11111111100000011011 | -0.502496 |
| | | | 11111111010011011011 | -0.4974910 |
| | | | 11111111011100100111 | -0.497491 |
| | | | 11111111010011010111 | 7.662695e-09 |
| USCI simulation with p=0.1 | -92.074847 | 40 | 11011111100111101011 | -0.999338 |
| | | | 11011111010100101000 | -0.036358 |
| | | | 00011111010100100100 | -2.039769e-08 |
| | | | 00011111100111101011 | 2.017419e-08 |
| | | | 11011111100111101000 | -1.468569e-08 |

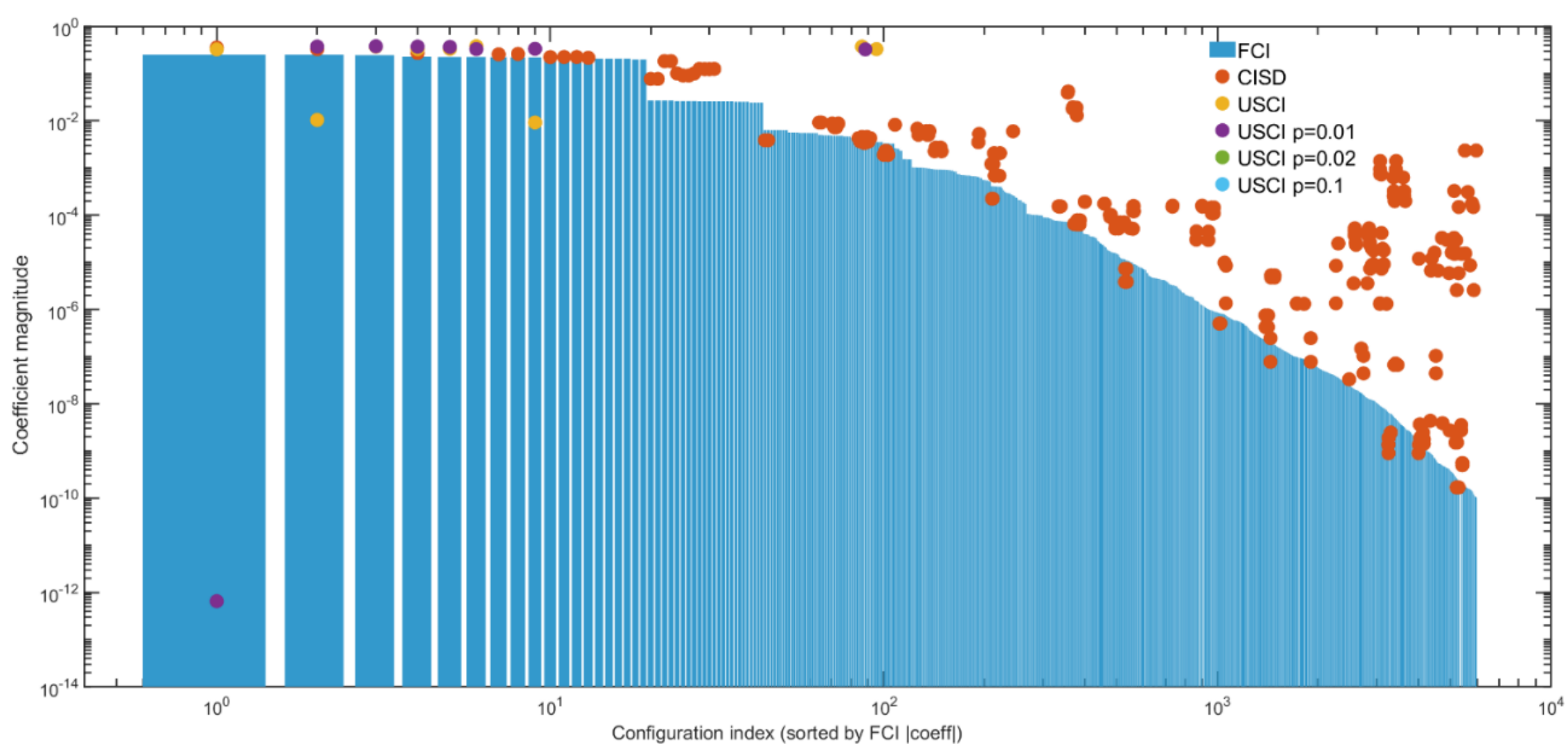

**Fig. S4 Determinant-amplitude spectrum for the 20-qubit $N_2$ (3.0 Å) benchmark.** Blue bars show the FCI coefficient magnitudes in descending order. Colored markers indicate the determinants retained by CISD, noiseless USCI, and USCI under depolarizing noise with $p$=0.01, 0.02, and 0.10. Mild noise leaves the dominant sparse structure largely intact, whereas stronger noise drives sampling into progressively smaller-amplitude regions of the determinant spectrum.

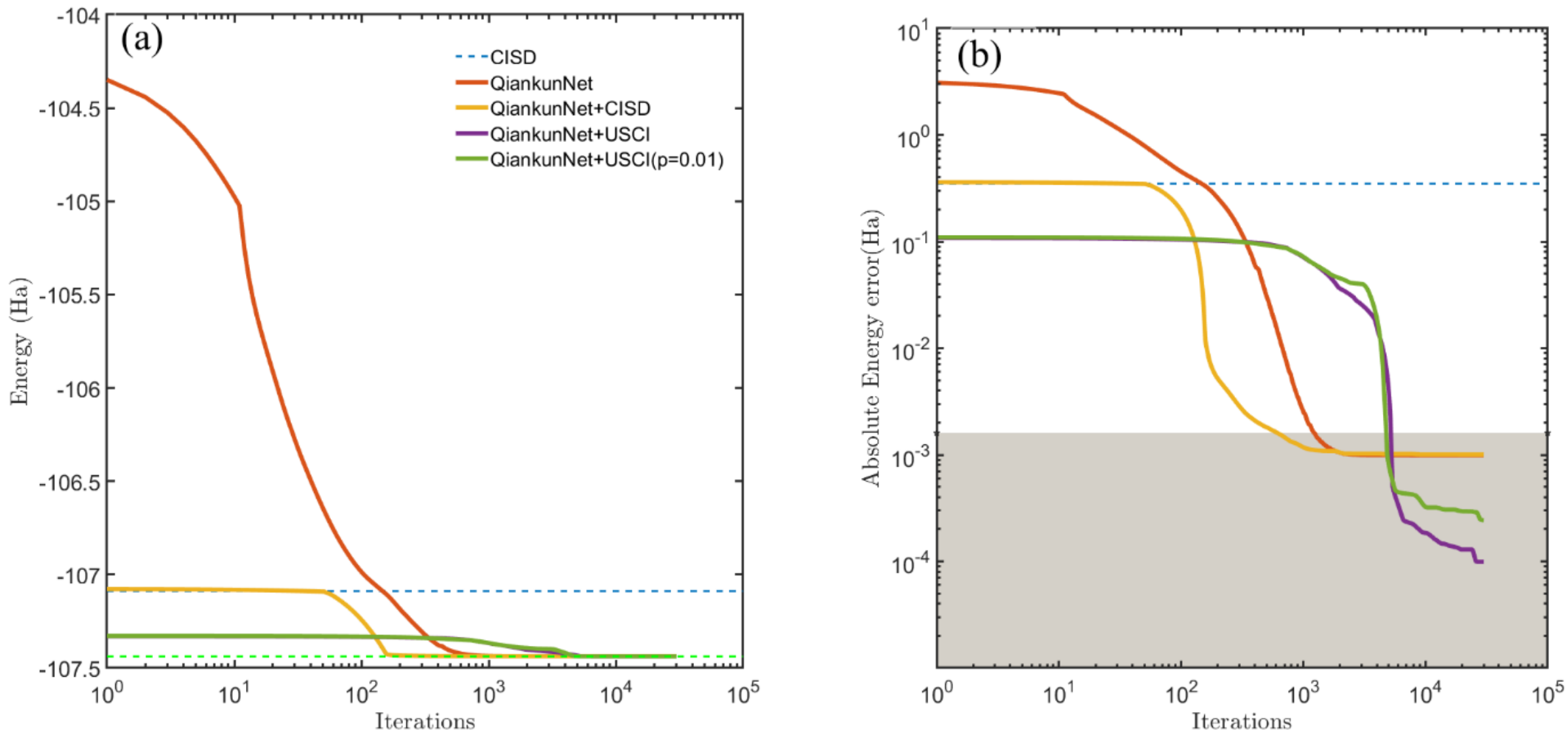

**Fig. S5 Effect of USCI-based initialization on QiankunNet optimization for the 20-qubit $N_2$ (3.0 Å) benchmark.** (a) Total energy versus optimization iteration for stand-alone QiankunNet and for hybrid variants initialized from CISD, noiseless USCI, and noisy USCI determinant pools. (b) Corresponding absolute energy error on a logarithmic scale. The shaded band indicates the chemical-accuracy window.

Fig. S5 addresses the more important question for the full workflow: whether a determinant pool generated by focused quantum sampling is useful even when it is not exact. The answer is yes, provided the sampling bias remains moderate. QiankunNet pretrained from the USCI-derived determinant pool converges more rapidly than the randomly initialized baseline, because the quantum step already concentrates probability on a compact sector of configuration space that contains much of the chemically relevant structure.

## 3.3 H-Couple expansion for quantum-sampled determinant spaces

The determinant pool generated by USCI-guided quantum sampling is not assumed to be complete. To recover additional correlation beyond the sampled determinant manifold while remaining variational, we enlarge the sampled subspace by explicitly adding determinants connected to it through the electronic Hamiltonian and then re-diagonalize in the augmented space. This procedure defines the Hamiltonian-coupling (H-Couple) expansion[14].

We work in an orthonormal Slater determinant basis built from spin orbitals, with the nonrelativistic electronic Hamiltonian written in second quantization:

$$H = \sum_{pq} h_{pq} a_p^\dagger a_q + \frac{1}{2} \sum_{pqrs} g_{pqrs}\, a_p^\dagger a_q^\dagger a_s a_r \tag{19}$$

where $h_{pq}$ and $g_{pqrs}$ are one and two-electron integrals respectively. According to the Slater–Condon rules, $\langle D'|H|D\rangle$ is nonzero only if $|D'\rangle$ differs from $|D\rangle$ by single or double excitations.

By the Slater–Condon rules, Hamiltonian matrix elements are non-zero only between determinants differing by one or two orbital substitutions. Let $S$ denote the determinant subspace obtained from USCI-guided quantum sampling after de-duplication and symmetry filtering. Diagonalization in $S$ yields a seed state:

$$|\Psi_S\rangle = \sum_{I \in S} c_I\, |I\rangle, \qquad E_S = \frac{\langle \Psi_S|H|\Psi_S\rangle}{\langle \Psi_S|\Psi_S\rangle} \tag{20}$$

with $E_S$ an upper bound to the FCI ground-state energy. We then define the Hamiltonian-connected

external set:

$$\mathcal{N}(S) = \{|\mu\rangle \notin S \,|\, \exists\, I \in S \text{ with } H_{\mu I} = \langle \mu|H|I\rangle \neq 0\} \tag{21}$$

Each $|\mu\rangle \in \mathcal{N}(S)$ is reachable from some $|I\rangle \in S$ by one or two orbital substitutions. Each determinant in this external set is reachable from the retained subspace by a single or double excitation. To prioritize the most chemically relevant couplings, we assign each candidate an amplitude-weighted score based on both its Hamiltonian coupling to $S$ and the amplitudes already present in the seed wavefunction,

$$s_\mu = \sum_{I \in S} |H_{\mu I} c_I| \tag{22}$$

with threshold $\tau$ (or retain the top-K by $s_\mu$). This score favors determinants that are both strongly coupled to $S$ and supported by the current amplitudes $\{\{c_I\}$.

$$C_\tau = \{|\mu\rangle \in \mathcal{N}(S): s_\mu \geq \tau\} \tag{23}$$

form the augmented space $S' = S \cup C_\tau$. Because the determinant basis is orthonormal, we solve the standard eigenproblem:

$$H_{S'S'} c' = E' c' \tag{24}$$

where $H_{S'S'} = [\langle A|H|B\rangle]_{A,B \in S'}$. The new ground-state estimate $|\Psi_{S'}\rangle = \sum_{A \in S'} c'_A |A\rangle$ satisfies the variational inequality $E' \leq E_S$. Because the determinant basis is orthonormal, this is a standard Hermitian eigenvalue problem, and the resulting ground-state estimate is variationally bounded from above. In practice we solve it with a Davidson algorithm.

It is useful to distinguish H-Couple from perturbative corrections such as Epstein–Nesbet PT2 [15]. PT2 estimates an energy correction without modifying the variational subspace, whereas H-Couple explicitly augments that subspace and re-diagonalizes the Hamiltonian

$$\Delta E_{PT2} = -\sum_{\mu \notin S} \frac{\left|\sum_{I \in S} H_{\mu I} c_I\right|^2}{E_\mu - E_S} \tag{25}$$

H-Couple is thus non-perturbative and variational by construction. As τ→0, or equivalently as *K* is increased and the procedure iterated, the enlarged space approaches the full CI manifold and the resulting energy decreases monotonically toward the exact limit.

## 4. Unified resource-error model for focused quantum sampling on NISQ hardware

The accuracy of focused quantum sampling in QSCI is governed by three coupled factors: (i) the exact wavefunction weight retained in the selected determinant subspace, (ii) the redistribution of measurement probabilities caused by circuit noise, and (iii) the finite-shot uncertainty in identifying the important determinants. The important distinction is that the noisy measurement histogram is not itself the final energy estimator. In QSCI, quantum measurements are used to define a determinant subspace, while the final energy is obtained by classical diagonalization of the Hamiltonian projected into that subspace. Therefore, noise and finite sampling mainly affect the quality and stability of determinant selection, rather than producing a direct statistical energy estimator as in VQE-like measurements. This distinction is central to the robustness of QSCI-type methods.

Let the active-space Hamiltonian be $\hat{H}$ and let its exact ground state be expanded in the computational determinant basis as

$$|\Psi_0\rangle = \sum_{i=1}^{d} c_i\,|D_i\rangle, \;\; q_i = |c_i|^2, \qquad \sum_{i=1}^{d} q_i = 1\,, \qquad d = 2^N \tag{26}$$

where $N$ is the number of qubits, $|D_i\rangle$ denotes a spin-orbital occupation string, and determinant basis state, and $q_i$ is the exact determinant probability in the ground-state wavefunction. For a retained determinant set $\mathcal{S}_R$ of size $\mathcal{R}$, we define the exact retained ground-state weight as:

$$Q_R = \sum_{i\,\in\,\mathcal{S}_R} q_i \tag{27}$$

The projector onto the retained subspace is,

$$\hat{P}_R = \sum_{i\,\in\,\mathcal{S}_R} |D_i\rangle\langle D_i| \tag{28}$$

and the normalized projection of the exact ground state,

$$|\Psi_R\rangle = \frac{\hat{P}_R|\Psi_0\rangle}{\sqrt{Q_R}}, \qquad Q_R > 0 \tag{29}$$

The quantity $Q_R$ , rather than the noisy measured probability, is the central object controlling the deterministic subspace-truncation error.

### 4.1 Deterministic subspace-truncation error

The Rayleigh-Ritz energy in the retained subspace is:

$$E_R = \min_{|\phi\rangle\in span(\mathcal{S}_R)} \frac{\langle\phi|\hat{H}|\phi\rangle}{\langle\phi|\phi\rangle} \tag{30}$$

By the variational principle,

$$E_0 \leq E_R \leq \langle\Psi_R|\hat{H}|\Psi_R\rangle \tag{31}$$

For any two normalized state $|\psi\rangle$ and $|\phi\rangle$ with $\delta = |\phi\rangle - |\psi\rangle$, and for any real shift $\mu$ using the Cauchy-Schwarz inequality

$$\left|\langle\phi|\hat{H}|\phi\rangle - \langle\psi|\hat{H}|\psi\rangle\right| = \left|\langle\phi|\hat{H} - \mu\hat{I}|\phi\rangle - \langle\psi|\hat{H} - \mu\hat{I}|\psi\rangle\right|$$

$$= \left|\langle\phi|\hat{H} - \mu\hat{I}|\delta\rangle + \langle\delta|\hat{H} - \mu\hat{I}|\psi\rangle\right| \leq 2\left\|\hat{H} - \mu I\right\|\|\delta\| = 2\left\|\hat{H} - \mu I\right\|\||\phi\rangle - |\psi\rangle\| \tag{32}$$

Minimizing over $\mu$ gives the shift-invariant spectral half-width

$$\Lambda_H = \min_{\mu \in \mathbb{R}} \left\| \hat{H} - \mu I \right\| = \frac{E_{max} - E_{min}}{2} \tag{33}$$

Where $E_{max}$ and $E_{min}$ are the largest and smallest eigenvalues of $\hat{H}$. In practice, one may use the looser bound $\Lambda_H \leq \left\| \hat{H} \right\|$. Since

$$\langle \Psi_0 | \Psi_R \rangle = \sqrt{Q_R} \tag{34}$$

We have

$$\| |\Psi_R\rangle - |\Psi_0\rangle \|^2 = 2 - 2\sqrt{Q_R} \tag{35}$$

Therefore

$$0 \leq E_R - E_0 \leq 2\Lambda_H \sqrt{2 - 2\sqrt{Q_R}} \tag{36}$$

The formula is consisted with previous results[16]. Combining this with the spectral-width bound $E_R - E_0 \leq E_{max} - E_{min} = 2\Lambda_H$, a slightly tighter and globally valid form is

$$0 \leq E_R - E_0 \leq min \left\{ 2\Lambda_H, 2\Lambda_H \sqrt{2 - 2\sqrt{Q_R}} \right\} \tag{37}$$

When the missing weight $\eta_R = 1 - Q_R$ is small,

$$2 - 2\sqrt{Q_R} = 2 - 2\sqrt{1 - \eta_R} = \eta_R + O({\eta_R}^2) \tag{38}$$

and Eq. (36) reduces to the asymptotic estimate

$$E_R - E_0 \leq 2\Lambda_H \sqrt{\eta_R} \tag{38}$$

This bound formalizes the principle of focused determinant sampling: a compact determinant set can yield a small variational error if it retains most of the exact ground-state wavefunction weight.

**4.2 Gate-noise-induced probability redistribution**

Let $p_i^{id}$ be the ideal output probability of determinant $|D_i\rangle$ from the noiseless quantum circuit. Under a global depolarizing approximation,

$$D_p(\rho) = (1 - p)\rho + p\frac{\hat{I}}{d} \ , \ \mathrm{d} = 2^N \tag{39}$$

the noisy measured determinant probability becomes:

$$\tilde{P}_i = (1 - p)p_i^{id} + \frac{p}{d} \tag{40}$$

If each two-qubit gate contributes an effective depolarizing strength $p_g$, then the total effective depolarizing probability is:

$$p = 1 - \left(1 - p_g\right)^{n_{2q}} \approx n_{2q} p_g, \quad p_g \ll 1 \tag{41}$$

where $n_{2q}$ is the number of two-qubit gates. If the reported two-qubit fidelity $F_{2q}$ is used only as a circuit empirical survival fidelity, one may set $p_g \approx 1 - F_{2q}$. For a determinant set $\mathcal{S}_R$, the noisy cumulative measured probability is

$$\tilde{P}_R = \sum_{i \in \mathcal{S}_R} \tilde{p}_i = (1 - p)P_R^{id} + \frac{pR}{d} \tag{42}$$

Where $P_R^{id} = \sum_{i \in \mathcal{S}_R} p_i^{id}$. Only in the idealized analytical limit where the noiseless circuit samples the exact ground-state distribution $p_i^{id} = q_i$, do we have $P_R^{id} = Q_R$, and therefore

$$\tilde{P}_R = (1-p)Q_R + \frac{pR}{d} \tag{43}$$

In that special limit,

$$Q_R = \frac{\tilde{P}_R - \frac{pR}{d}}{1-p} \tag{44}$$

In realistic USCI sampling, $p_i^{id}$ is expected to be correlated with $q_i$, but it is not identical to $q_i$. It is therefore useful to define a circuit-distribution mismatch

$$\zeta_R = \left| P_R^{id} - Q_R \right| \tag{45}$$

Then a noise-corrected cumulative probability can only bound the exact retained ground-state weight up to this mismatch:

$$Q_R = \frac{\tilde{P}_R - \frac{pR}{d}}{1-p} - \zeta_R \tag{46}$$

Equation (46) makes explicit that the deterministic Rayleigh-Ritz error depends on exact retained weight $Q_R$, whereas the measure $\tilde{P}_R$ is only a noisy proxy used to infer or select the determinant subspace.

The global depolarizing model preserves the ordering of determinant probabilities in the infinite-shot limit, because Eq. (40) rescales all ideal probabilities by the same factor and adds the same uniform background. Thus, under strict global-depolarizing and infinite sampling, noise does not change the identity of the top-R determinants. Its main effect is to reduce probability contrast, which makes finite-shot determinant selection more difficult.

### 4.3 Finite-shot uncertainty for a fixed determinant set

Let *M* be the number of measurement shots. For a fixed determinant set $\mathcal{S}_R$, define

$$Y_R = \sum_{m=1}^{M} 1[x_m \in \mathcal{S}_R] \tag{47}$$

Where $x_m$ is the determinant measured in shot m. Then $Y_R \sim \text{Binomial}(M, \tilde{P}_R)$ and the empirical cumulative probability is $\hat{\tilde{P}}_R = \frac{Y_R}{M}$. Thus

$$\mathbb{E}\left[\hat{\tilde{P}}_R\right] = \tilde{P}_R \tag{48}$$

and

$$\text{Var}\left[\hat{\tilde{P}}_R\right] = \frac{\tilde{P}_R(1-\tilde{P}_R)}{M} = \frac{\frac{1}{4} - \left(\tilde{P}_R - \frac{1}{2}\right)^2}{M} \leq \frac{1}{4M} \tag{49}$$

A high-probability version follows from Hoeffding's inequality[17]:

$$\Pr\left[\left|\hat{\tilde{P}}_R - \tilde{P}_R\right| \geq \epsilon_M\right] \leq 2e^{-2M\epsilon_M^2} \tag{50}$$

Therefore, with confidence at least $1-\delta$,

$$\left|\hat{\tilde{P}}_R - \tilde{P}_R\right| \leq \epsilon_M(\delta), \qquad \epsilon_M(\delta) = \sqrt{\frac{\log\left(\frac{2}{\delta}\right)}{2M}} \tag{51}$$

Combining Eqn.(46) and (51), we obtain a lower confidence bound for the exact retained weight:

$$Q_R \geq Q_R^L = max\left[0, \frac{\hat{\tilde{P}}_R - \epsilon_M(\delta) - \frac{pR}{d}}{1-p} - \zeta_R\right] \tag{52}$$

Substituting this lower bound into the deterministic truncation estimate gives:

$$0 \leq E_R - E_0 \leq min\left\{2\Lambda_H, 2\Lambda_H\left(2 - 2\sqrt{Q_R^L}\right)^{\frac{1}{2}}\right\} \tag{53}$$

with confidence at least $1-\delta$, within the assumed depolarizing-noise model and the assumed circuit-distribution mismatch $\zeta_R$.

This formulation is preferable to adding a separate $\sqrt{R/M}$ energy term. In QSCI, measured probabilities are not directly used to compute the final Rayleigh-Ritz energy once the subspace has been selected. Finite shots mainly affect the reliability with which the determinant subspace is identified.

The previous subsection assumes a fixed determinant set $\mathcal{S}_R$. In practice, $\mathcal{S}_R$ is often selected from measured samples. Let the noisy probabilities within a candidate pool of size $K$ be sorted as

$$\tilde{p}_{(1)} \geq \tilde{p}_{(2)} \geq \cdots \geq \tilde{p}_{(K)} \tag{54}$$

Define the probability gap at the top-R boundary as $\Delta_R = \tilde{p}_R - \tilde{p}_{R+1}$. If every empirical probability satisfies:

$$\left|\hat{\tilde{p}}_i - \tilde{p}_i\right| < \frac{\Delta_R}{2} \tag{55}$$

then the empirical top-R set is identical to the true noisy top-R set. By Hoeffding's inequality and a union bound over the *K* candidate determinants,

$$\Pr\left[\hat{\mathcal{S}}_R \neq \mathcal{S}_R\right] \leq 2K\exp\left(-\frac{M\Delta_R^2}{2}\right) \tag{56}$$

Under the global depolarizing model,

$$\Delta_R = (1-p)\left(p_{(R)}^{id} - p_{(R+1)}^{id}\right) \tag{57}$$

Therefore, recovering the same top-R determinant set with probability at least 1−δ satisfies

$$M \geq \frac{2\log\left(\frac{2K}{\delta}\right)}{(1-p)^2\left(p_{(R)}^{id} - p_{(R+1)}^{id}\right)^2} \tag{58}$$

Equation (59) shows the central resource coupling: gate noise reduces the probability gap by the factor $1-p$, and the required number of measurements grows as $(1-p)^{-2}$.

If the top-R set is recovered correctly, the final Rayleigh-Ritz error is controlled by Eq. (37). If the selection fails, a conservative expectation-level bound is

$$\mathbb{E}\left[E_{\hat{\mathcal{S}}_R} - E_0\right] \leq min\left\{2\Lambda_H, 2\Lambda_H\left(2 - 2\sqrt{Q_R}\right)^{\frac{1}{2}}\right\} + 2\Lambda_H Pr\left[\hat{\mathcal{S}}_R \neq \mathcal{S}_R\right] \tag{59}$$

Combining Eqs. (56) and (57) gives the scaling estimate

$$\mathbb{E}\left[E_{\hat{\mathcal{S}}_R} - E_0\right] \leq min\left\{2\Lambda_H, 2\Lambda_H\left(2 - 2\sqrt{Q_R}\right)^{\frac{1}{2}}\right\} + 4K\Lambda_H\exp\left[-\frac{M(1-p)^2\left(p_{(R)}^{id} - p_{(R+1)}^{id}\right)^2}{2}\right] \tag{60}$$

This expression is mathematically cleaner than an additive $\sqrt{R/M}$ energy term. It shows that finite-shot effects enter QSCI primarily through determinant-selection failure, and that this failure probability is jointly controlled by shot number, circuit noise, and the probability gap at the selection boundary.

For comparison, if one directly estimates the energy of the noisy prepared state rather than performing QSCI subspace diagonalization, the depolarizing channel produces a direct energy bias. For

$$\rho_{noise} = (1-p)\rho_{id} + p\frac{\hat{I}}{d} \tag{61}$$

The energy shit is:

$$\Delta E_{noise} = \left|Tr\left[\hat{H}\rho_{noise}\right] - Tr\left[\hat{H}\rho_{id}\right]\right| = p\left|\frac{Tr\left[\hat{H}\right]}{d} - Tr\left[\hat{H}\rho_{id}\right]\right| \tag{62}$$

Since both $\frac{Tr[\hat{H}]}{d}$ and $Tr\left[\hat{H}\rho_{id}\right]$ lie within the spectral interval $[E_{min}, E_{max}]$, a worst-case bound is

$$\Delta E_{noise} \leq 2p\Lambda_H \tag{63}$$

This term is relevant for direct VQE-like energy estimation, but it should not be confused with the QSCI truncation error after classical diagonalization. In QSCI, noisy measurements mainly affect which determinants are selected, while the Hamiltonian matrix used in the final diagonalization is evaluated classically.

### 4.4 Uniform sampling versus structured quantum sampling

For a closed-shell active space CAS(n,m), where n is the number of spatial orbitals and m is the number of electrons, the number of spin-resolved particle-conserving determinants is

$$\Omega_{CAS} = \binom{n}{m/2}^2 \tag{64}$$

The corresponding qubit register contains 2n spin orbitals, so the total number of computational-basis bit strings is $2^{2n}$. Therefore, the probability that a uniformly random bit string lies in the correct closed-shell particle-number sector is

$$P_\mu = \frac{\binom{n}{m/2}^2}{2^{2n}} \tag{65}$$

**For CAS(10o,10e),** $P_\mu = \frac{\binom{10}{5}^2}{2^{20}} \approx 0.0606$. If a structured quantum circuit has $N_g$ two-qubit gates, each with effective survival fidelity $F_{2q}$, then the no-error trajectory probability is approximately $P_{surv} \approx F_{2q}^{N_g}$. Requiring the no-error circuit-survival probability to exceed the uniform probability of landing in the valid determinant sector gives:

$$F_{2q}^{N_g} > P_u \tag{66}$$

Since $0 < F_{2q} < 1$, this implies:

$$N_g < \frac{\log P_u}{\log F_{2q}} \tag{67}$$

Substituting Eq. (64), we obtain:

$$N_g < \frac{2\log\binom{n}{m/2} - 2n\log 2}{\log F_{2q}} \tag{68}$$

For CAS(10o,10e), Eq. (67) gives approximately $N_g < 279$ for $F_{2q}$=0.990 and $N_g < 699$ for $F_{2q}$=0.996. This comparison should be interpreted only as an order-of-magnitude illustration. $P_{surv}$ is not the probability of sampling a particular determinant; it is the approximate probability that the circuit output is not strongly corrupted by two-qubit gate errors. The actual advantage of structured quantum sampling comes from concentrating probability on chemically important determinants, not merely from satisfying particle-number constraints.

## 5. QiankunNet architecture, variational optimization, and computational settings

### 5.1 Wavefunction parameterization and training objective

QiankunNet serves as the classical refinement module in the QiankunNet-QSCI workflow. After determinant-focused quantum sampling with the USCI ansatz, the sampled determinant set and its associated coefficients are passed to QiankunNet, which learns a compact variational representation of the many-electron wavefunction and performs classical reweighting and completion of the sampled subspace. In the present implementation, the wavefunction is factorized into amplitude and phase components,

$$\Psi_\Theta(x) = A_{\Theta_A}(x) \exp\left(i\phi_{\Theta_\phi}(x)\right) \tag{69}$$

where $x$ denotes the occupation string of a Slater determinant, $A_{\Theta_A}(x)$ is the amplitude network, and $\phi_{\Theta_\phi}(x)$ is the phase network. This amplitude–phase decomposition allows QiankunNet to represent a complex-valued wavefunction while keeping the learning problem structurally aligned with the determinant basis used in QSCI.

Internally, QiankunNet adopts a decoder-only transformer backbone that processes determinant occupation strings as input sequences and uses masked multi-head self-attention to capture long-range orbital correlations. This design is particularly suitable for strongly correlated systems, where important correlation patterns are often nonlocal in orbital space and cannot be captured reliably by local heuristics alone. Starting from the determinant subset generated by the quantum sampling stage, the network parameters are optimized variationally by minimizing the energy expectation value in the selected determinant subspace. In practice, the Hamiltonian expectation value is evaluated over the retained determinant manifold, gradients are backpropagated through the neural network, and the model parameters are updated iteratively until convergence. In this way, QiankunNet does not merely fit the raw sampled histogram; rather, it corrects sampling bias, redistributes weight toward chemically important but under-sampled determinants, and improves the final variational estimate of the target state.

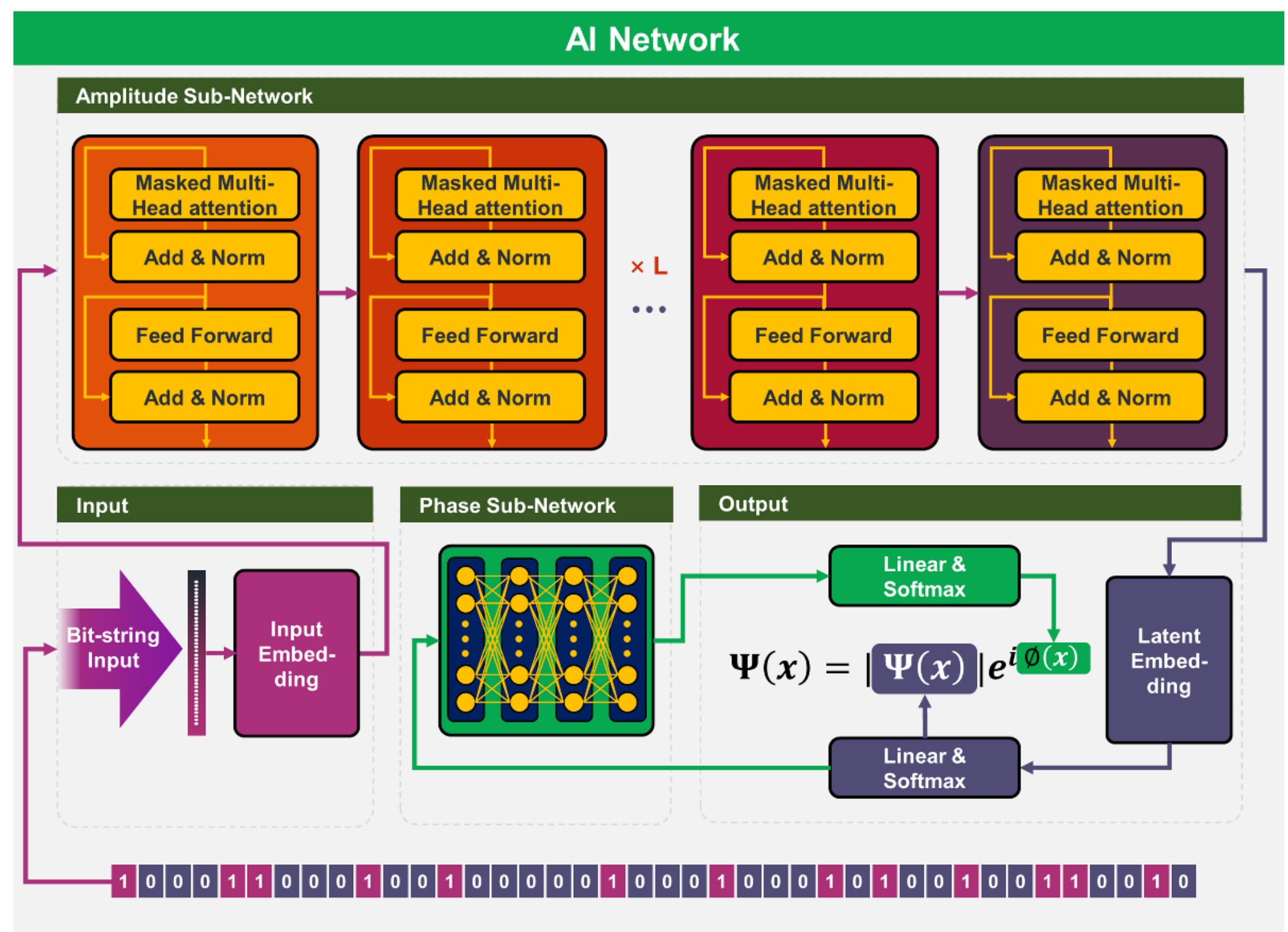

**Fig. S6 Schematic of the QiankunNet architecture with separate amplitude and phase subnetworks.**

## 5.2 Model hyperparameters for [2Fe-2S] and P-cluster

The practical hyperparameters used in the reported calculations are summarized in Table S5. For the [2Fe-2S] calculation, the transformer backbone uses a model dimension of 32, four transformer layers, and four attention heads. The associated MLP readout has layer widths [40,512,512,512,512,1], giving a total of 861,313 trainable parameters. For the P-cluster calculation, the model size is increased to a model dimension of 96 with the same number of transformer layers but six attention heads; the corresponding MLP readout is [146,512,512,512,512,1], with 1,319,265 trainable parameters in total. The first dimension of the readout MLP matches the spin-orbital input size of each target system, namely 40 for [2Fe-2S] and 146 for the P-cluster.

For both systems, the neural network is optimized using the AdamW optimizer with a learning rate of $3\times10^{-4}$ and momentum parameters $\beta_1$=0.9 and $\beta_2$=0.99. Inference is performed in batches of 8192 determinants for [2Fe-2S] and 2048 determinants for the P-cluster. The larger model dimension and reduced batch size used for the P-cluster reflect the substantially larger active space and the increased memory demand associated with determinant scoring, variational reweighting, and downstream subspace refinement.

## 5.3 GPU hardware and computational cost

All QiankunNet training and inference calculations reported in this work were carried out on NVIDIA H100 GPUs. The total computational cost for the [2Fe-2S] calculation was 562.6 GPU-hours, while the corresponding cost for the P-cluster calculation was approximately 7,219.2 GPU-hours. The marked increase in cost from [2Fe-2S] to the P-cluster is consistent with the much larger active space, the more complex determinant connectivity, and the heavier workload associated with large-scale neural-wavefunction refinement in the P-cluster calculation.

Taken together, these implementation details clarify the practical role of QiankunNet in the overall workflow. The quantum hardware is responsible for producing a determinant distribution enriched in chemically important configurations, whereas QiankunNet provides the classical variational machinery needed to denoise, reweight, and complete that sampled determinant manifold. This division of labor is central to the scalability of the QiankunNet-QSCI framework, especially for large strongly correlated systems such as the P-cluster, where direct brute-force quantum sampling or purely classical determinant exploration would both become prohibitively inefficient.

**Table S5. QiankunNet hyperparameters and computational cost for the [2Fe-2S] and P-cluster calculations**.

| System | Input size | D_model | Transformer layers | Attention heads | MLP widths | Trainable parameters | Inference batch size | Optimizer | Learning rate | Betas | GPU hardware | Total GPU-hours |
|---|---|---|---|---|---|---|---|---|---|---|---|---|
| [2Fe-2S] | 40 | 32 | 4 | 4 | [40, 512, 512, 512, 512, 1] | 861,313 | 8,192 | AdamW | 3e-4 | (0.9, 0.99) | NVIDIA H100 | 562.6 |
| P-cluster | 146 | 96 | 4 | 6 | [146, 512, 512, 512, 512, 1] | 1,319,265 | 2,048 | AdamW | 3e-4 | (0.9, 0.99) | NVIDIA H100 | ~7,219.2 |

# 6. Quantum experimental results and QiankunNet-QSCI simulations

## 6.1 $H_{10}$: hardware validation of determinant-focused sampling

We examine the $H_{10}$ ring in an active space of 10 electrons in 10 orbitals as a controlled test case for hardware-assisted determinant focusing. This system is large enough to exhibit a nontrivial determinant distribution, yet small enough that full configuration interaction (FCI) remains available as an exact benchmark.

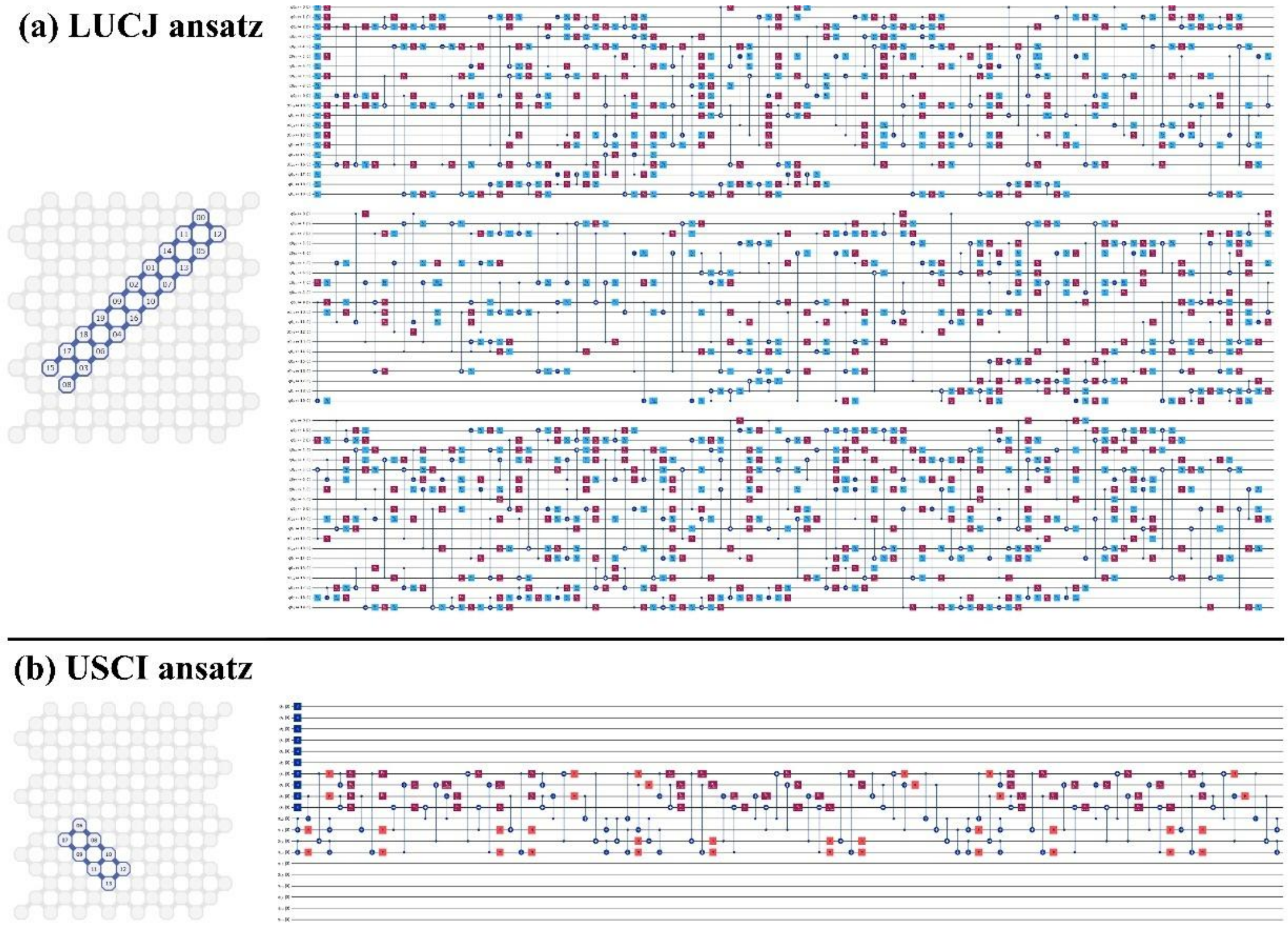

**Fig. S7 Hardware-compiled ansatz for the $H_{10}$ benchmark on *Zuchongzhi* 3.1.** (a) LUCJ circuit compiled for the full 20-qubit (10e,10o) active-space representation. Left: qubit layout of the 20-qubit square lattice with the entangler path highlighted. Right: scheduled native-gate circuit after hardware compilation. (b) USCI circuit compiled on an 8-qubit problem-adaptive support. For this benchmark, the dominant selected excitations involve only 8 spin orbitals, so the compiled USCI circuit acts non-trivially on an 8-qubit sub-register, while the remaining orbitals are kept in their Hartree–Fock occupations. Left: selected qubit sub-lattice. Right: scheduled native-gate circuit.

For $H_{10}$, the full electronic problem is defined in the 10e,10o active space, corresponding to 20 spin orbitals. However, after determinant prescreening, the dominant selected excitations entering the USCI ansatz are confined to only 8 spin orbitals. We therefore compile the non-trivial USCI block on an 8-qubit sub-register, while keeping the remaining spin orbitals frozen in their Hartree–Fock occupations during the hardware sampling stage. The sampled bit strings are then re-embedded into the full 20-qubit occupation representation before determinant filtering and QiankunNet initialization. Figure S7 illustrates the hardware-level distinction between the two circuit constructions. LUCJ distributes variational freedom over a comparatively dense entangling structure, whereas USCI uses a more selective circuit derived from a compact prescreened determinant manifold.

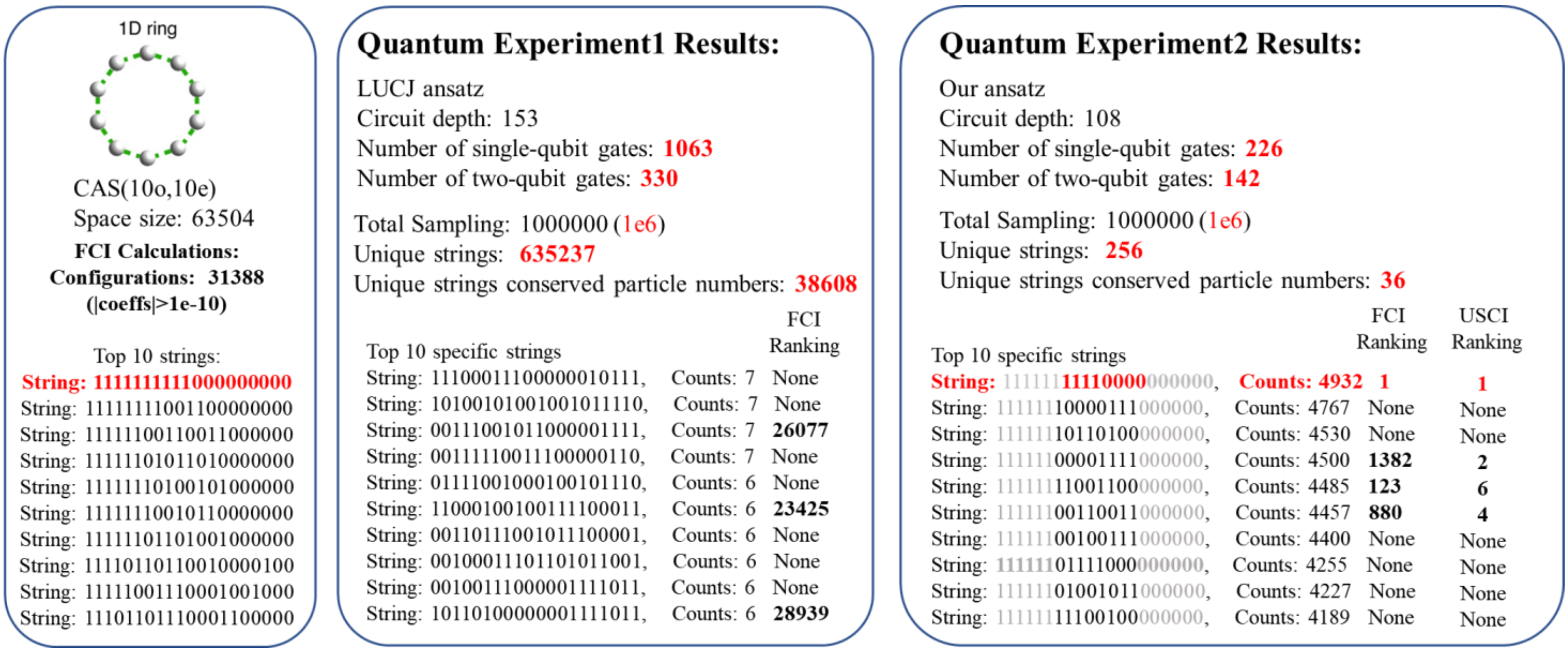

**Fig. S8 Quantum sampling distributions for the $H_{10}$ ring in the (10e,10o) active space.** Left: dominant determinants from the full 20-qubit FCI reference. Middle: hardware samples obtained from the LUCJ circuit acting on the full 20-qubit register. Right: hardware samples obtained from the USCI circuit, whose non-trivial support is restricted to 8 qubits for this benchmark and subsequently re-embedded into the full 20-qubit occupation representation for comparison.

Fig. S8 shows that the two ansatzes produce qualitatively different sampling behavior. The LUCJ circuit yields a broad distribution over a very large number of measured bit strings, only a fraction of which satisfy the target particle-number constraint. In contrast, the USCI circuit generates a much more concentrated set of outputs with substantially fewer unique strings, while still recovering physically relevant determinants, including dominant configurations identified from FCI. The point of this comparison is not that LUCJ is intrinsically ineffective as a variational ansatz, but that for the sampling task relevant here, USCI is markedly more selective: it places probability mass on a much smaller and chemically more useful region of Hilbert space.

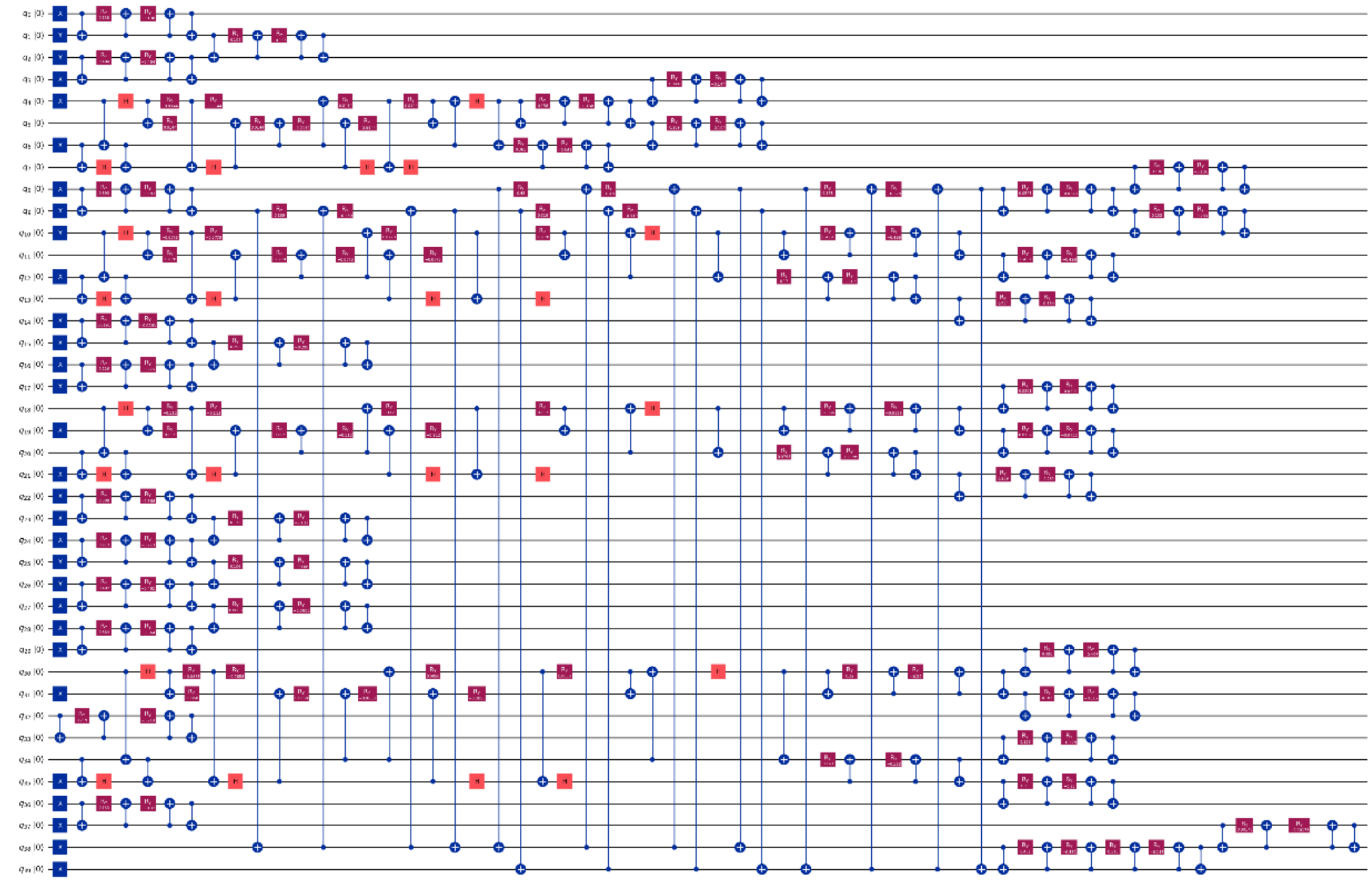

**Fig. S9 Hardware-compiled USCI ansatz for the [2Fe-2S] cluster on *Zuchongzhi* 3.1.**

The consequences of this difference become clearer when the sampled determinant pools are passed to QiankunNet. Figs. S10-S13 track the evolution of determinant-weight distributions during classical refinement, comparing stand-alone QiankunNet with versions initialized from LUCJ- and USCI-derived hardware samples.

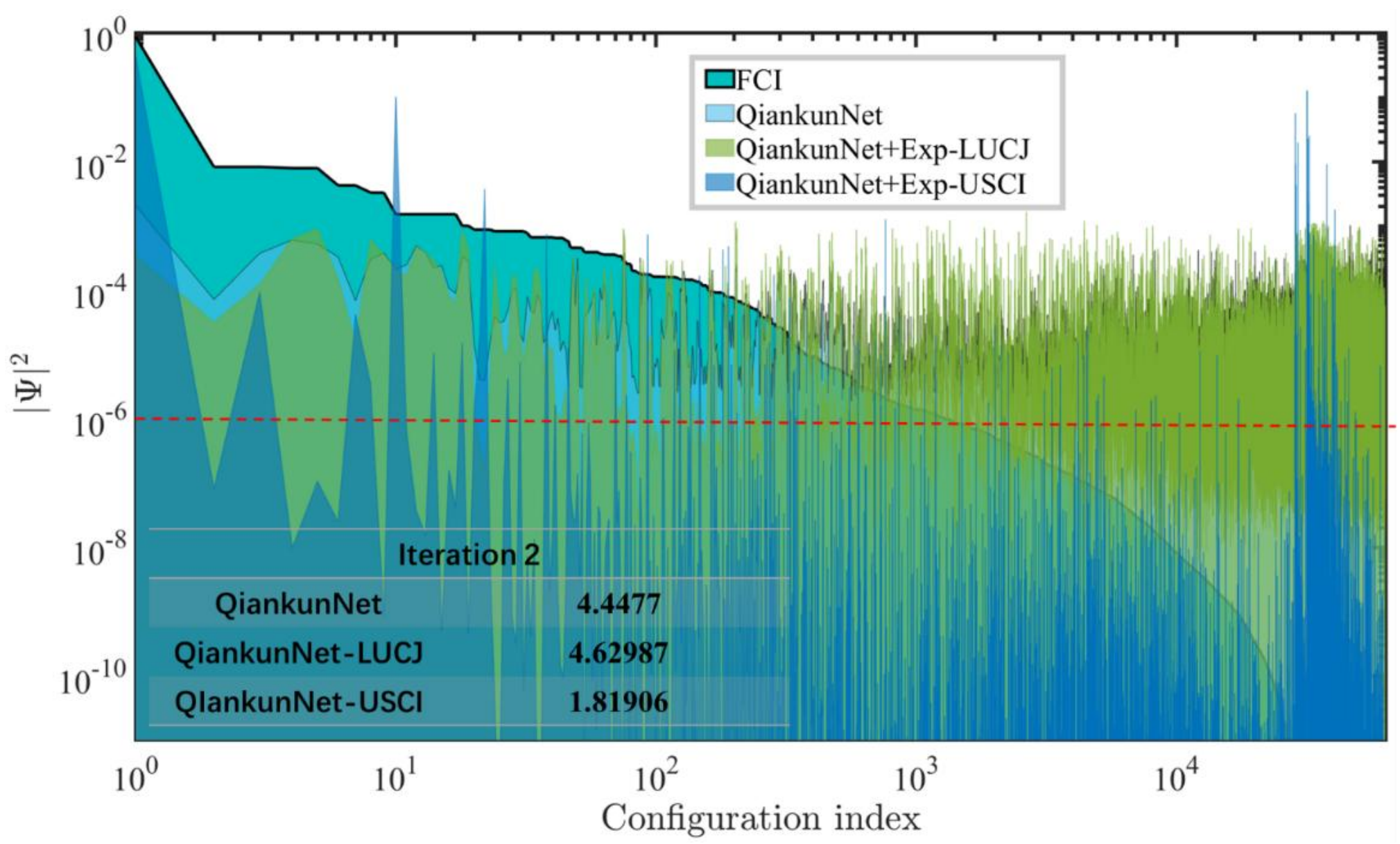

**Fig. S10 Determinant-weight distributions after 2 optimization iterations for the $H_{10}$ benchmark.** Squared determinant coefficients are shown for QiankunNet alone, QiankunNet initialized from LUCJ hardware samples, and QiankunNet initialized from USCI hardware samples, together with the FCI reference distribution. The dashed horizontal line marks the determinant-selection threshold used in the analysis.

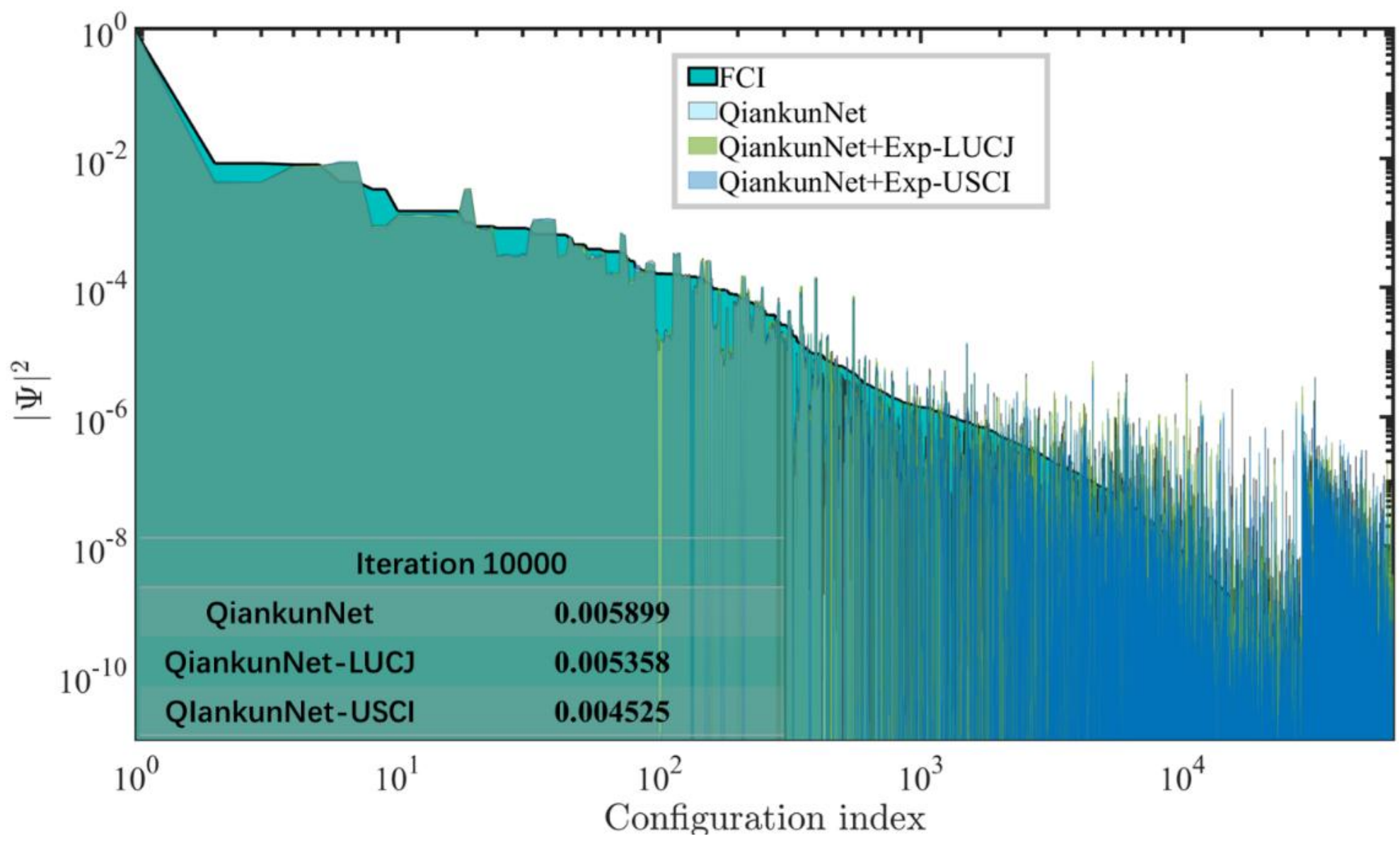

**Fig. S11 Determinant-weight distributions after 10,000 optimization iterations for the $H_{10}$ benchmark.** Same format as Fig. S10.

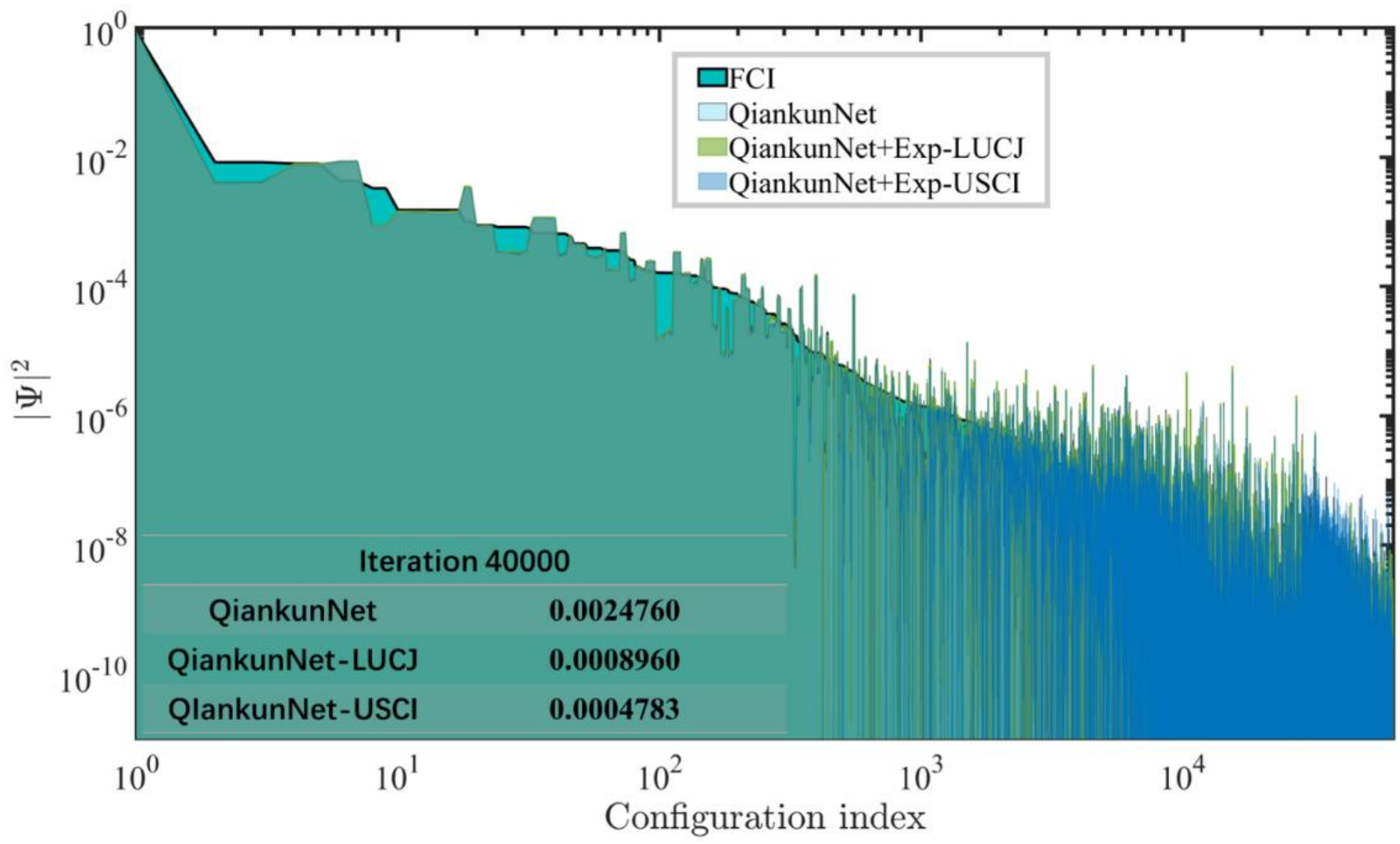

**Fig. S12 Determinant-weight distributions after 40,000 optimization iterations for the $H_{10}$ benchmark.** Same format as Fig. S10.

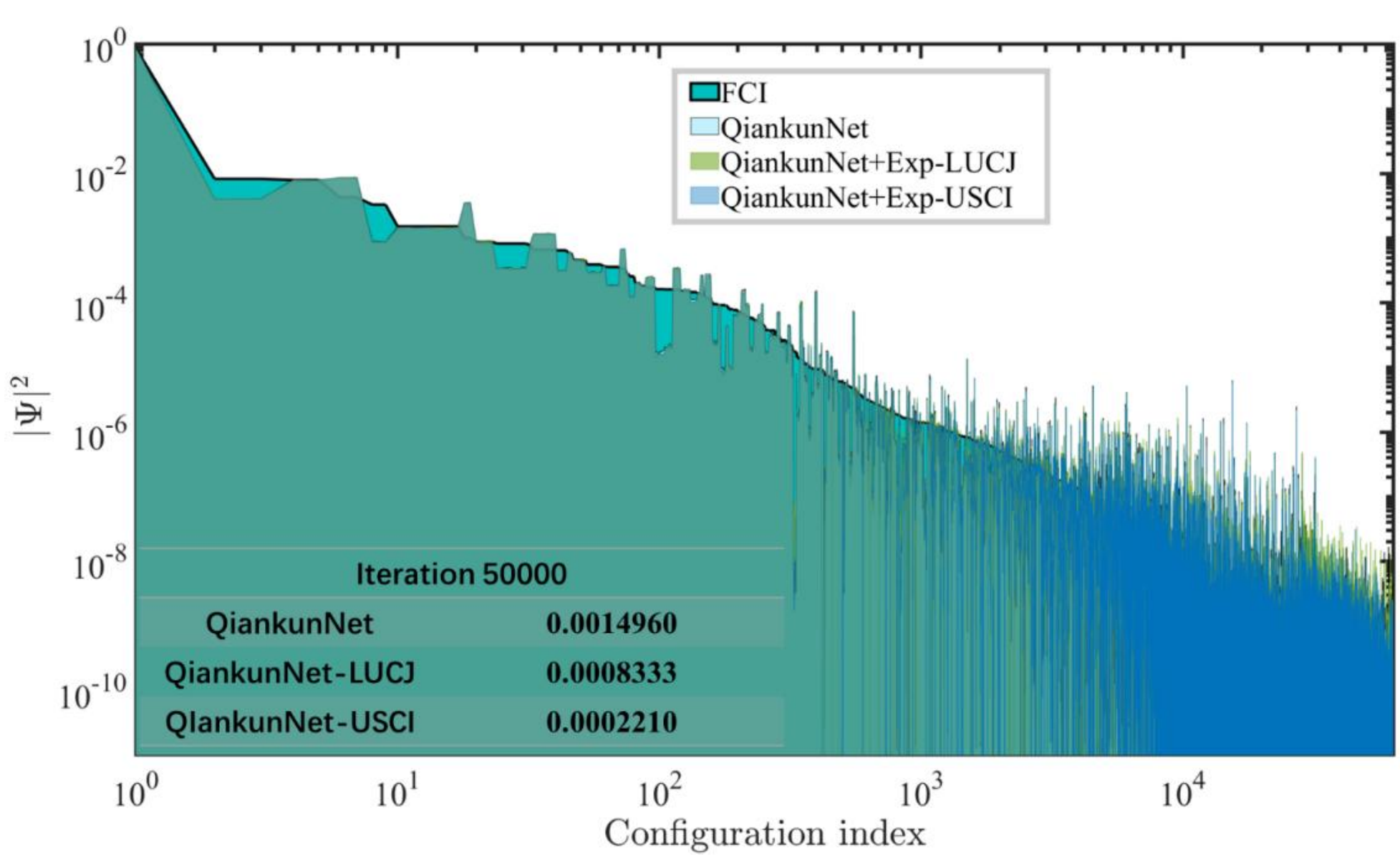

**Fig. S13 Determinant-weight distributions after 50,000 optimization iterations for the $H_{10}$ benchmark.** Same format as Fig. S10.

Taken together, Figs. S10-S13 show that the advantage of USCI is not limited to the raw hardware samples themselves. When used to initialize QiankunNet, the USCI-derived determinant pool leads to a more rapid reshaping of the learned coefficient distribution toward the FCI reference than either the uninitialized baseline or the LUCJ-seeded variant. At very early optimization stages this difference is already visible, and at later stages it remains reflected in the smaller residual energy of the USCI-seeded workflow.

### 6.2 [2Fe-2S]: from focused sampling to chemically meaningful convergence

The $H_{10}$ benchmark establishes that USCI can concentrate sampling on a chemically relevant determinant sector on real hardware. We next consider the 40-qubit $[Fe_2S_2(SCH_3)_4]^{2-}$ cluster, which is the main chemically relevant hardware-assisted target of the manuscript.

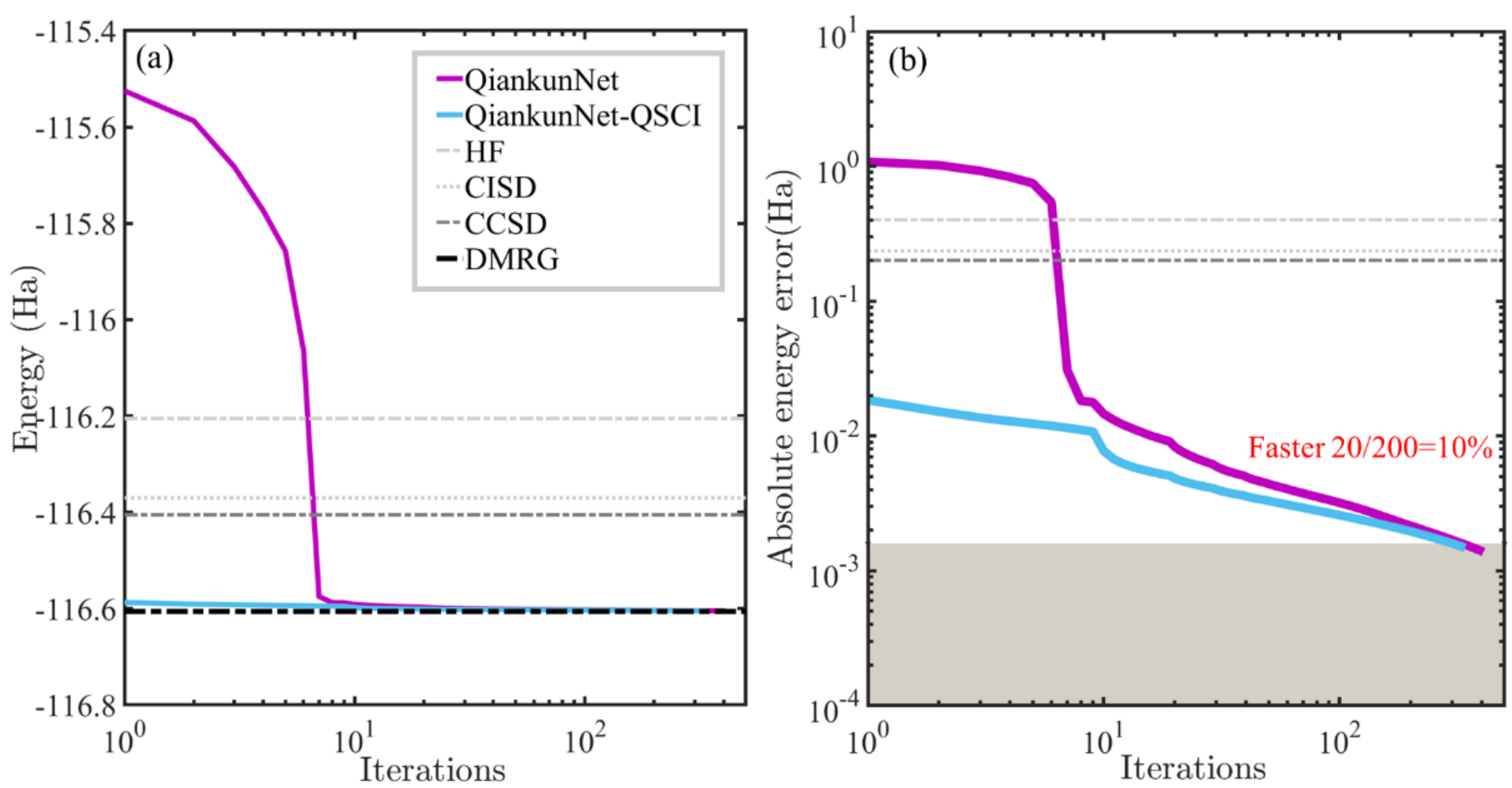

**Fig. S14 Convergence of QiankunNet and QiankunNet-QSCI for the [2Fe-2S] cluster.** (a) Total energy versus optimization iteration. Horizontal lines denote HF, CISD, CCSD, and DMRG reference energies. (b) Absolute error relative to DMRG. The shaded band denotes the 1 mHa chemical-accuracy window.

Figure S14 shows that QiankunNet-QSCI converges substantially more rapidly toward the DMRG reference than QiankunNet without quantum initialization. Figure S15 clarifies why focused sampling is particularly valuable for this system. The coefficient distribution is strongly heavy-tailed: although the formal FCI space is enormous, a comparatively small leading sector carries a disproportionate share of the norm, while most determinants contribute only tiny weights. This is precisely the regime in which determinant-focused sampling becomes algorithmically useful. If a shallow quantum circuit can preferentially enrich the leading chemically important sector, then subsequent learning and subspace diagonalization need not begin from an unstructured or nearly uniform determinant pool.

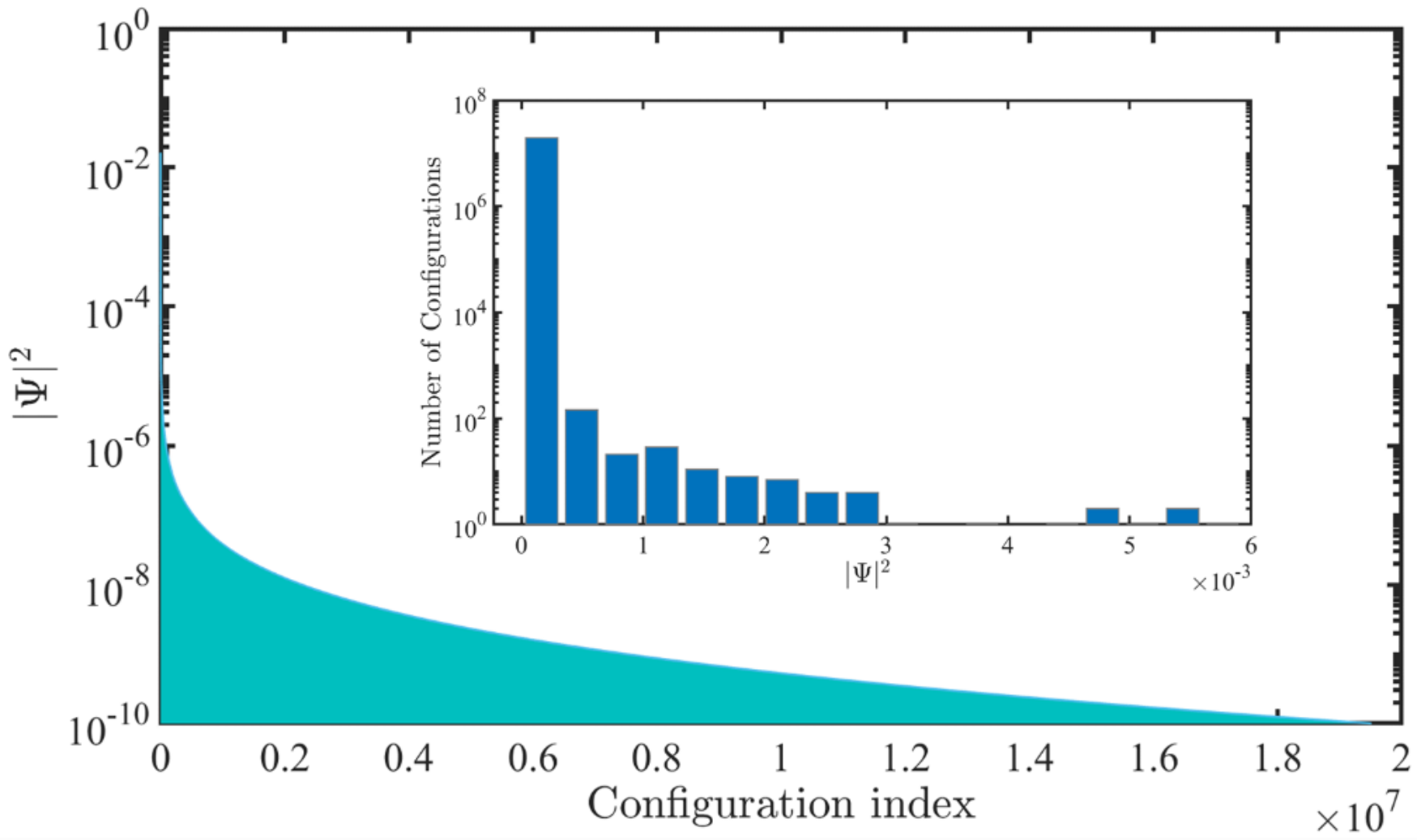

**Fig. S15 Full-CI coefficient landscape for the [2Fe–2S] cluster in the CAS(30e, 20o) active space.** Squared CI coefficients are shown in descending order for the spin-adapted determinant expansion. The inset reports the corresponding histogram of determinant weights. The distribution is strongly heavy-tailed, with a relatively narrow leading sector carrying a disproportionate fraction of the total wavefunction norm.

### 6.3 P-cluster: spin-factorized hardware sampling at large scale

The P-cluster provides the strongest scalability test in the present work. In this case, the central question is no longer whether a focused determinant pool improves refinement on a moderate-size benchmark, but whether hardware-assisted sampling can still identify a chemically meaningful subspace when the full unfactorized Hilbert space is far beyond the reach of direct brute-force exploration. The purpose of this subsection is therefore to show how the determinant-focusing strategy is adapted to a much larger active space and why spin factorization is essential to making the problem hardware compatible.

To map the CAS(114e, 73o) active space onto the available superconducting architecture, we exploit spin factorization and treat the α- and β-spin sectors separately. The 146 spin orbitals are thus partitioned into two 73-qubit registers, each implemented as an independent USCI circuit on *Zuchongzhi* 3.1. This is not a change in the underlying chemistry, but a hardware-oriented reorganization of the sampling problem: the α and β sectors are sampled independently at the circuit level, after which the resulting determinant information is recombined during the classical refinement stage.

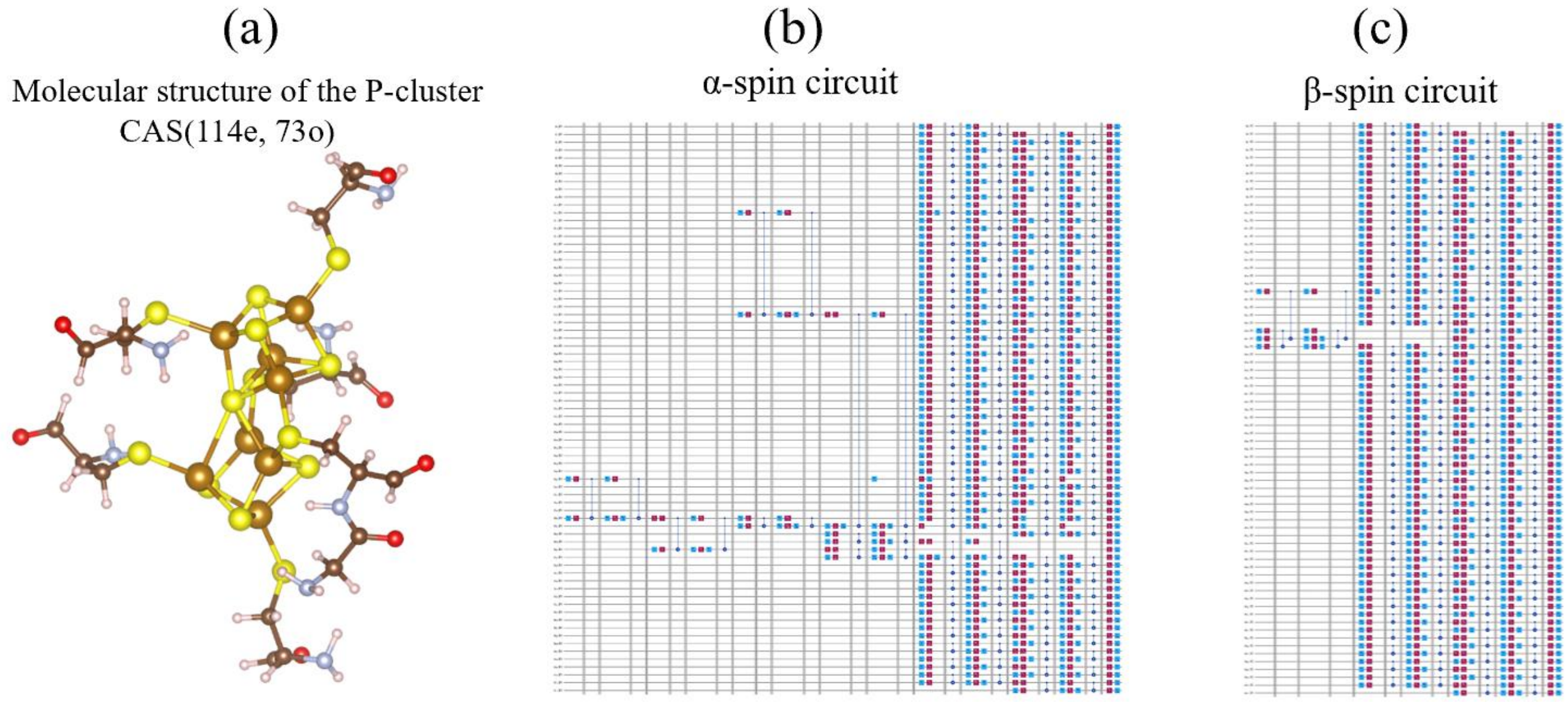

**Fig. S16 Spin-factorized USCI circuits for the P-cluster benchmark.** (a) Molecular structure of the nitrogenase P-cluster and the associated CAS(114e, 73o) active space. (b) Compiled α-spin USCI circuit on a 73-qubit register of *Zuchongzhi* 3.1. (c) Compiled β-spin USCI circuit on a second 73-qubit register. Reported gate counts and shot numbers summarize the hardware resources used for sampling in each spin sector.

The reference state for each spin sector is taken from the leading determinant of a preliminary SHCI calculation. Because the broken-symmetry magnetic structure distributes α-only and β-only occupations differently across the Fe centers, the two spin sectors have distinct hole patterns and therefore different excitation structures. This immediately leads to different compiled circuits for the αand β channels, as seen in Fig. S16. The purpose of showing both circuits is not merely to report hardware metrics, but to make clear that the spin-factorized construction is chemically informed: the circuit difference reflects the underlying spin asymmetry of the reference electronic structure, rather than an arbitrary compilation artifact. The practical advantage of this factorization can be quantified by comparing the observed hardware yield with a uniform-sampling baseline.

**Table S6: Quantum sampling efficiency relative to uniform sampling for the P-cluster benchmark.** For each spin sector, the table reports the number of particle-number-conserving strings obtained from $10^8$ measurement shots, the corresponding expectation for uniform random sampling over the same 73-qubit space, and the resulting enrichment factor.

| Sampling Metric (per $10^8$ shots) | Alpha (α) Space | Beta (β) Space |
|---|---|---|
| Total Measurement Shots | $10^8$ | $10^8$ |
| Valid Strings (Particle Conserving) | 1,345 | 3,335 |
| Expected Valid Strings (Uniform Sampling) | *≈56* | *≈56* |
| Quantum Importance Sampling Enhancement | 24× | 60× |

Note: The theoretical probability of selecting a valid particle-conserving configuration via uniform sampling in this specific space is strictly calculated as $C(73,57)/2^{73} \approx 5.6\times10^{-7}$. For $10^8$ total measurement shots, the mathematical expectation is ~56 valid physical configurations. The raw hardware yield of 1,345 and 3,335 valid strings is consistent with our USCI circuits performing physics-informed importance sampling, concentrating quantum probability amplitude into the chemically relevant sub-manifold despite NISQ-level device noise.

Table S6 shows that the USCI-guided circuits produce far more valid particle-number-conserving strings than would be expected from uniform random sampling over the same qubit space. The hardware is not merely generating large numbers of raw measurement shots, but is concentrating those shots into a physically relevant submanifold of the spin-factorized Hilbert space. The hardware provides enriched α- and β-sector determinant pools; these are then combined in the α⊗β tensor-product space, where QiankunNet refinement and H-Couple expansion recover the inter-spin correlation.

## 7. Wavefunction analysis of [2Fe-2S]

### 7.1 Spin-orbital entanglement entropy

We formulate the entropy and mutual-information measures at the spin-orbital level. For a given spin orbital *i*, we define the marginal occupation probability as the total probability of all determinants in which that spin orbital is occupied:

$$p_i = \sum_{\substack{k \\ D_k(i)=1}} p_k \tag{70}$$

where $D_k(i)$ denotes the *i*th bit of the determinant $D_k$. The corresponding single-spin-orbital entropy $s_i$ is then:

$$s_i = -[p_i ln p_i + (1 - p_i)\, ln(1 - p_i)] \tag{71}$$

This quantity measures how uncertain the occupation of spin orbital *i* remains within the reconstructed many-electron wavefunction. Small values indicate orbitals that are close to always empty or always occupied, whereas values approaching the maximum indicate spin orbitals that participate strongly in the correlated manifold.

### 7.2 Mutual information

For a pair of distinct spin orbitals (*i,j*), we define the joint occupation probabilities over the four binary occupation sectors $(a,b) \in \{0,1\}^2$:

$$p_{ab}^{(i,j)} = \sum_{\substack{k \\ D_k(i)=a \\ D_k(j)=b}} p_k \tag{72}$$

From this distribution we obtain the joint entropy

$$s_{ij} = -\sum_{a,b\in\{0,1\}} p_{ab}^{(i,j)}\, ln\left(p_{ab}^{(i,j)}\right) \tag{73}$$

and the mutual information $I(i,j)$

$$I(i,j) = s_i + s_j - s_{ij} \tag{74}$$

The mutual information measures the total correlation between the two spin orbitals, including both classical correlation and quantum entanglement contributions.[18-19] Larger values therefore identify pairs of spin orbitals whose occupations are strongly coupled in the correlated wavefunction.

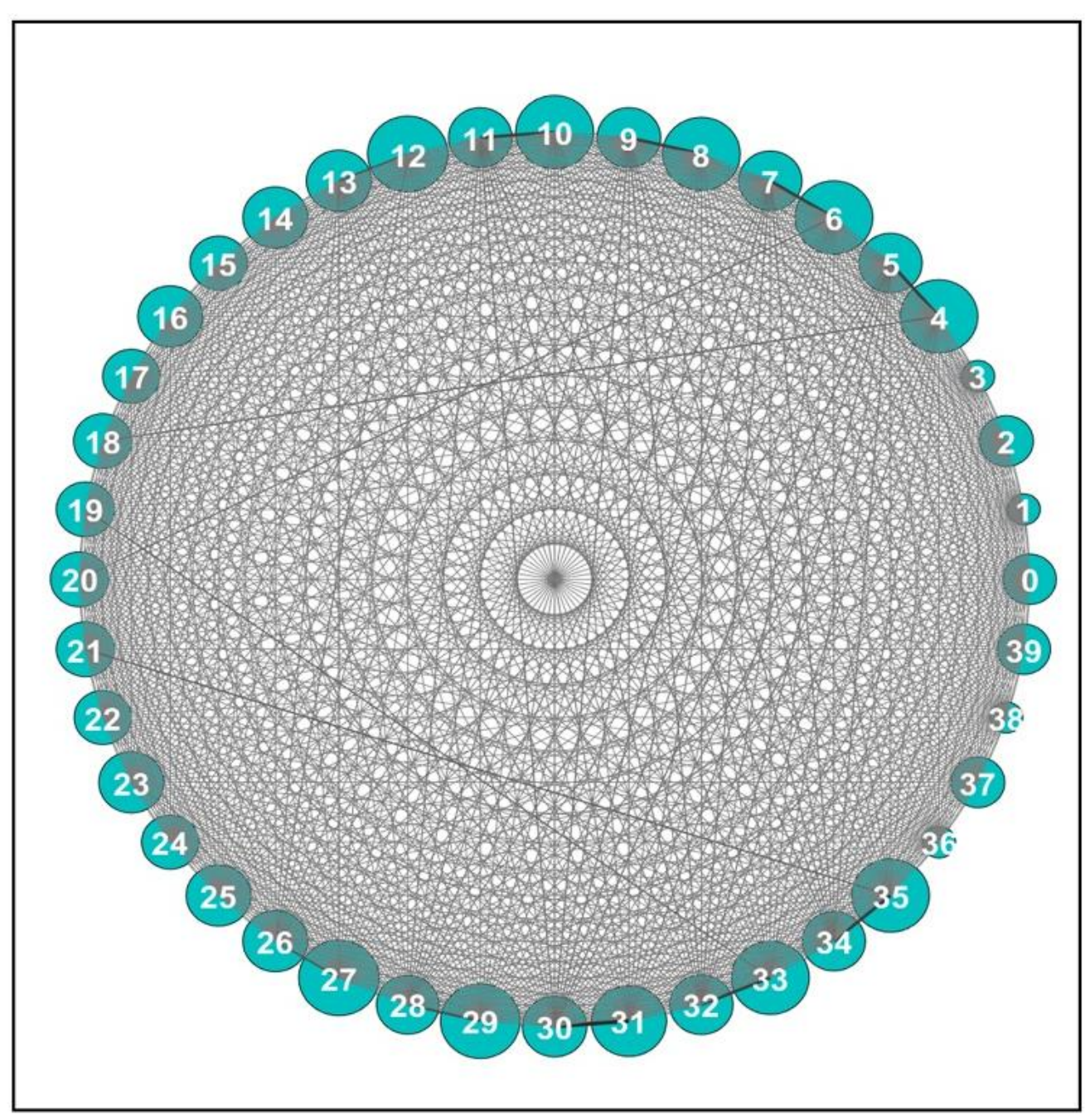

**Fig. S17 Mutual-information network for the [2Fe–2S] cluster in the CAS(30e, 20o) active space**. The node size proportional to single-orbital entropy and edge thickness proportional to pairwise mutual information.